\documentclass[12pt,preprint]{aastex}
\usepackage{epsfig}
\usepackage{natbib}          
\shorttitle{CHANDRA OBSERVATIONS OF NGC~4365 \& NGC~4382 (M85)}
\shortauthors{SIVAKOFF, SARAZIN, \& IRWIN}
\slugcomment{Astrophysical Journal, submitted}
\slugcomment{Draft version of \today}

\received{2002 December 13}
\begin{document}

\title{\textit{Chandra} Observations of Low Mass X-ray Binaries and Diffuse
Gas in the Early-Type Galaxies NGC~4365 and NGC~4382 (M85)}

\author{Gregory R. Sivakoff\altaffilmark{1},
Craig L. Sarazin\altaffilmark{1}, and
Jimmy A. Irwin\altaffilmark{2}}

\altaffiltext{1}{Department of Astronomy, University of Virginia,
P. O. Box 3818, Charlottesville, VA 22903-0818;
grs8g@virginia.edu, sarazin@virginia.edu}

\altaffiltext{2}{Department of Astronomy, University of Michigan,
Ann Arbor, MI 48109-1090; jirwin@astro.lsa.umich.edu}

\begin{abstract}
We used the \textit{Chandra X-ray Observatory} ACIS S3 to image the X-ray
faint elliptical galaxy NGC~4365 and lenticular galaxy NGC~4382. The
observations resolve much of the X-ray emission into 99 and 58 sources,
respectively, most of which are low-mass X-ray binaries (LMXBs) associated
with each of the galaxies.
Within one effective radius of NGC~4365, about 45\% of the counts are resolved
into sources, 30\% are attributed to unresolved LMXBs, and 25\% are attributed
to diffuse gas. Within two effective radii of NGC~4382, about 22\% of the
counts are resolved into sources, 33\% are attributed to unresolved LMXBs,
and 45\% are attributed to diffuse gas.
We identify 18 out of the 37 X-ray sources in a central field in NGC~4365
with globular clusters.
The luminosity functions of the resolved sources for both galaxies are best
fit with cutoff power-laws whose cutoff luminosity is
$\approx 0.9 - 3.1 \times 10^{39}$ ergs s$^{-1}$. These luminosities are much
larger than those previously measured for similar galaxies; we do not find
evidence for a break in the luminosity function at the Eddington luminosity of
a 1.4 $M_\odot$ neutron star.
The spatial distributions of the resolved sources for both galaxies are broader
than the distribution of optical stars.
In both galaxies, a hard power-law model fits the summed spectrum
of all of the sources.
The unresolved emission is best fit by the sum of a soft mekal model
representing emission from diffuse gas, and a hard power-law,
presumed to be from unresolved LMXBs.
There is some evidence that the temperature of the diffuse gas increases
with increasing radius.
A standard beta model fits the radial distribution of the diffuse
gas in both galaxies.
In the elliptical NGC~4365, the best-fit core radius is very small, while the
S0 galaxy NGC~4382 has a larger core radius. This may indicate that the gas in
NGC~4382 is rotating significantly.
\end{abstract}


\section{Introduction} \label{sec:intro}

Early-type galaxies are luminous X-ray sources \citep*{FJT1985}. For galaxies
of a given optical luminosity, the X-ray-to-optical luminosity ratio
($L_X/L_B$) ranges over two orders of magnitude for galaxies
\citep*{CFT1987,WD1997}. We will refer to galaxies with relatively high
$L_X/L_B$ ratios as ``X-ray bright'' and to galaxies with relatively low
$L_X/L_B$ ratios as ``X-ray faint.'' Hot ($kT \sim 1$ keV) interstellar gas
dominates the X-ray emission in X-ray bright galaxies (e.g., \citealt{FJT1985};
\citealt*{TFC1986}); whereas X-ray faint galaxies exhibit two distinct
spectral components: a hard ($\sim$5--10 keV) component \citep{MKA+1997} and a
very soft ($\sim$ 0.2 keV) component (\citealt*{FKT1994}; \citealt{P1994};
\citealt{KFM+1996}). Since the hard component is actually found in both X-ray
bright and X-ray faint early-type galaxies, with strengths roughly proportional
to the optical luminosity of the galaxy, \citet{KFT1992} suggested that the
hard component is due to low-mass X-ray binaries (LMXBs) like those observed
in the Milky Way's bulge.

\textit{Chandra} observations of the X-ray faint elliptical, NGC~4697
\citep*{SIB2000,SIB2001}, resolved the majority of emission into X-ray point
sources, whose properties are consistent with LMXBs. Chandra observations show
that a significant fraction of the LMXBs are coincident with globular clusters
\citep{SIB2000,ALM2001}. Taken together, the X-ray spectra of LMXBs in
early-type galaxies are hard, and can be fit by thermal bremsstrahlung with a
temperature of $\sim 7$ keV. However, there is a considerable variety in
observed X-ray spectra and colors of individual sources. Some of the LMXBs
have very soft X-ray spectra, similar to those of Galactic supersoft sources
\citep{SIB2000}.
The luminosity function of LMXBs in early-type galaxies in the luminosity range
$5 \times 10^{37} - 10^{39}$ erg s$^{-1}$ appears to generally be fit by a
broken power-law \citep{SIB2000,SIB2001}, with a break luminosity which is
comparable to the Eddington luminosity for spherical accretion onto a 1.4
$M_\odot$ neutron star. This suggests that the sources more luminosity than
$\sim 2 \times 10^{38}$ erg s$^{-1}$ are accreting black holes.

In addition to the LMXBs, the \textit{Chandra} observations of NGC~4697 also
detected diffuse emission from interstellar gas.
The gas is quite cool, with temperatures $\sim 0.3$ keV
\citep{SIB2000,SIB2001}.

With \textit{Chandra}'s ability to study both the LMXB population and diffuse
X-ray emission of early-type galaxies, it is useful to extend such studies to
other early-type galaxies. Since diffuse emission could dominate the LMXBs in
X-ray bright galaxies, the ideal targets for studying the LMXB population of
early-type galaxies are X-ray faint galaxies. In this paper, we present the
results of two such observations. Both NGC~4365 and NGC~4382 are early-type
galaxies located in the Virgo cluster. NGC~4365 is an E3 galaxy. NGC~4382 (M85)
is a S0 galaxy with an interacting companion NGC~4394, which together with
several other nearby galaxies (VCC~797, IC~3292) form a galaxy group RSCG 54
\citep{BGR+1996}. Since NGC~4382 has somewhat bluer colors than are typical for
a S0, and its disk shows a spiral pattern, it is sometimes classified as an Sa.
It was the host of the Type Ia supernova SN1960r \citep[e.g.,][]{P1993}. The
distances to these galaxies are 20.4 Mpc (NGC~4365) and 18.5 Mpc (NGC~4382),
based on the method of surface brightness fluctuations \citep{TDB+2001}. For
comparison, NGC~4697 has a distance of 11.7 Mpc using the same technique. All
three galaxies have similar X-ray luminosities and $L_X/L_B$ ratios
\citep*{OFP2001}.

In \S~\ref{sec:obs}, we discuss the observations and data reduction of these
galaxies. The X-ray images are presented in \S~\ref{sec:image}. The properties
of resolved sources are given and discussed in \S~\ref{sec:sources}. We
discuss the X-ray spectral properties in \S~\ref{sec:spectral_analysis}. The
spatial distribution of the diffuse X-ray emitting gas is determined in
\S~\ref{sec:dif_dist}. We summarize our conclusions in \S~\ref{sec:conclusion}.

\section{Observation and Data Reduction} \label{sec:obs}

NGC~4365 was observed on 2001 June 2--3 with a live exposure of 40,429 s,
while NGC~4382 was observed on 2001 May 29--30 for 39,749 s. Both galaxies
were observed with the ACIS-235678 chips operated at a temperature of
$-$120~C and with a frame time of 3.2~s. We determined the pointings so that
the entire galaxy was located on the S3 chip and so that the center of each
galaxy was not on a node boundary of the chip. Although a number of
serendipitous sources were seen on the other chips, the analysis of both
galaxies in this paper will be based on data from the S3 chip alone. The data
were telemetered in Faint mode, and only events with ASCA grades of 0,2,3,4,
and 6 were included. Photon energies were determined using the gain file
acisD2000-08-12gainN0003.fits. We excluded bad pixels, bad columns, and columns
adjacent to bad columns or chip node boundaries.

{\it Chandra} is known to encounter periods of high background (``background
flares''), which especially affect the backside-illuminated S1 and S3
chips\footnote{See \url{http://cxc.harvard.edu/contrib/maxim/acisbg/}.}.
We determined the background count rate for each galaxy, using the S1 chip to
avoid the enhanced flux due to the galaxy on the S3 chip. For both NGC~4365 and
NGC~4382, no background flares were seen. The backgrounds for extended regions
were taken from the deep blank sky backgrounds compiled by Maxim
Markevitch\footnotemark[3], and were adjusted to the aspect histories of our
observations using his {\sc make\_acisbg} program\footnotemark[3]
The normalization of the background was increased very slightly
to match the 5--10 keV count rate of the blank-sky background with a
galaxy emission-free region of each galaxy on the S3 chip.
We included the ``background'' due to the readout artifact in ACIS using his
{\sc make\_readout\_bg} program\footnotemark[3].

On the S3 image of NGC~4365, there were two X-ray sources with positions that
agreed with positions from the U.S. Naval Observatory A2.0
\citep[USNOA2][]{MCD+1998} or the Two Micron All Sky Survey
\citep[2MASS][]{CSV+2001} catalogs. Src.~98 in Table~\ref{tab:src_n4365}, an
X-ray source near the edge of the field where the {\it Chandra}
point-spread-function (PSF) is quite broad and the X-ray position is uncertain,
corresponded to USNOA2 0900-07143676. For this source, the optical and X-ray
positions agree to within $\sim$0\farcs4. The second possible optical
identification is between the optical nucleus of NGC~4365 itself and Src.~1 in
Table~\ref{tab:src_n4365}. Unfortunately, the central region of this galaxy is
crowded with sources, and it is uncertain whether the source closest to the
nucleus is, in fact, an active galactic nucleus (AGN), a nearly coincident
X-ray binary, or a combination of sources. Confusion limits the accuracy with
which the positions of sources near the center of NGC~4365 can be determined.
In the absence of more direct evidence, we will assume that the positions in
NGC~4365 are accurate to 1\arcsec. In NGC~4382, there were two X-ray sources
with optical identifications and positions from the Tycho-2 catalog
\citep{HFM+2000}, the 2MASS catalog, and/or the USNOA2 optical catalog. The
first source, Src.~38 in Table~\ref{tab:src_n4382}, corresponds to J1225+182,
a $z=1.19$ quasar \citep{KB1998} and is the brightest X-ray source in the S3
field. The second source, which was just below the detection threshold for
inclusion in Table~\ref{tab:src_n4382}, corresponds to BD+18 2609, a G5 star
with a proper motion of $\sim$90 mas~yr$^{-1}$ \citep{HKB1998}. The X-ray
positions agreed with the optical positions to better than 0\farcs4 in both
cases with no evidence for a systematic offset. We believe that the X-ray
positions for NGC~4382 are generally accurate to $\sim$0\farcs5.

We performed initial data reduction and some of the data analysis using the
{\it Chandra} analysis package {\sc ciao}\footnote{See
\url{http://asc.harvard.edu/ciao/}.}. We extracted spectra using the
{\sc calcrmf}\footnote{See \url{http://asc.harvard.edu/cont-soft/software/}.}
software kindly provided by Alexey Vikhlinin and Jonathon McDowell. The spectra
were fit using {\sc xspec}\footnote{See
\url{http://heasarc.gsfc.nasa.gov/docs/software/lheasoft/}.}.
Since launch, the ACIS quantum efficiency (QE) has undergone continuous
degradation\footnote{See
\url{http://cxc.harvard.edu/cal/Acis/Cal\_prods/qeDeg/}.}.
We used the {\sc xspec acisabs} model to correct X-ray spectra for this
effect.

\section{X-ray Image} \label{sec:image}

\begin{figure}
\plottwo{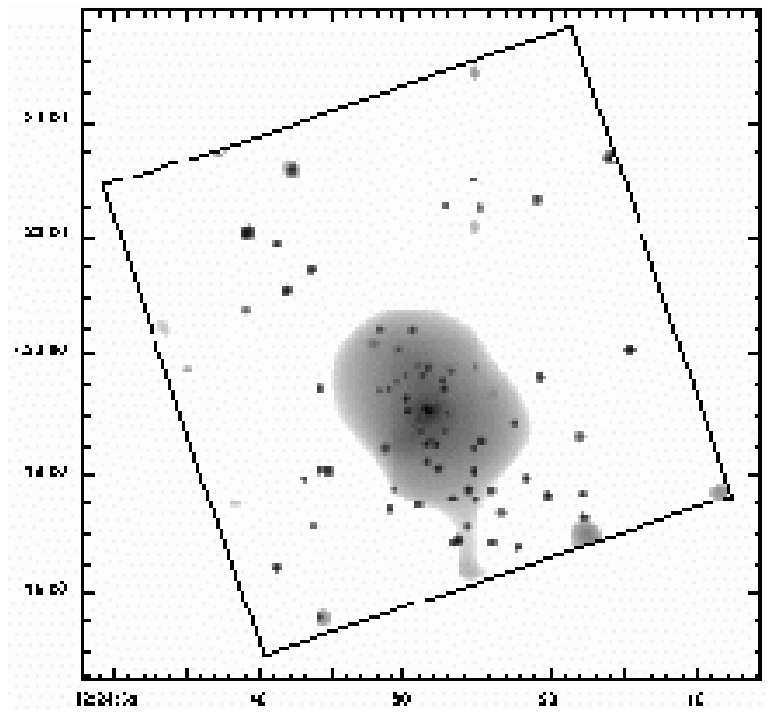}{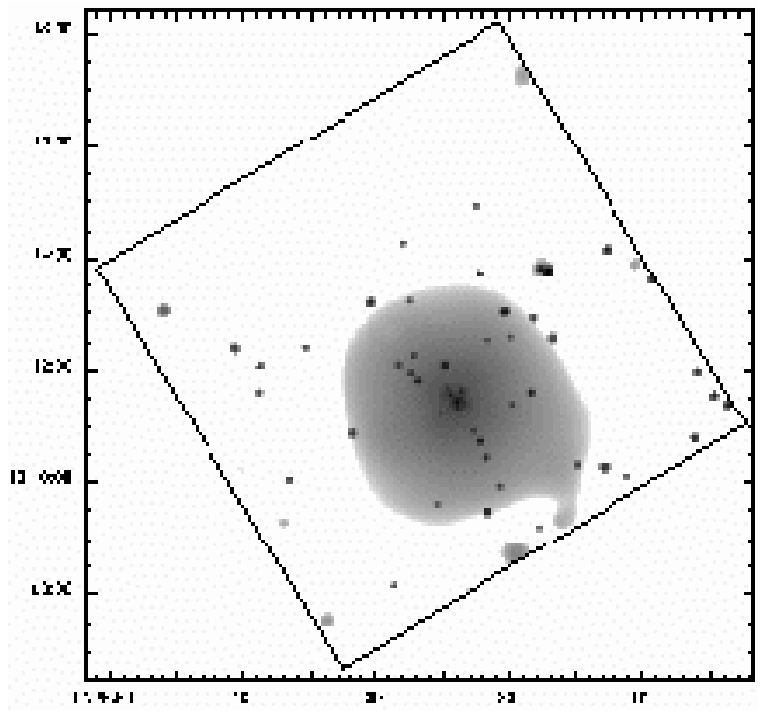}
\caption{
Adaptively smoothed {\it Chandra} S3 image (0.3--10 keV) of (left) NGC~4365 and
(right) NGC~4382, corrected for exposure and background. The gray scale is
logarithmic and ranges from $1.56 \times 10^{-6}$ cnt pix$^{-1}$ s$^{-1}$ to
$3.3 \times 10^{-4}$ cnt pix$^{-1}$ s$^{-1}$.
\label{fig:adaptive}}
\end{figure}

The X-ray images of NGC~4365 and NGC~4382 show both resolved point sources and
diffuse emission. To best image this combination, we adaptively smoothed the
{\it Chandra} S3 image using a minimum signal-to-noise ratio (SNR) per
smoothing beam of 3. After correcting for exposure and background, we removed
regions of less than 20 ks to avoid artifacts at the chip's edge.
Figure~\ref{fig:adaptive} displays gray scale images with a logarithmic
surface brightness scale ranging between
$1.56 \times 10^{-6}$ cnt pix$^{-1}$ s$^{-1}$ and
$3.3 \times 10^{-4}$ cnt pix$^{-1}$ s$^{-1}$.
In both galaxies, the majority of the X-ray luminosity emission has been
resolved into point sources.

\begin{figure}
\plottwo{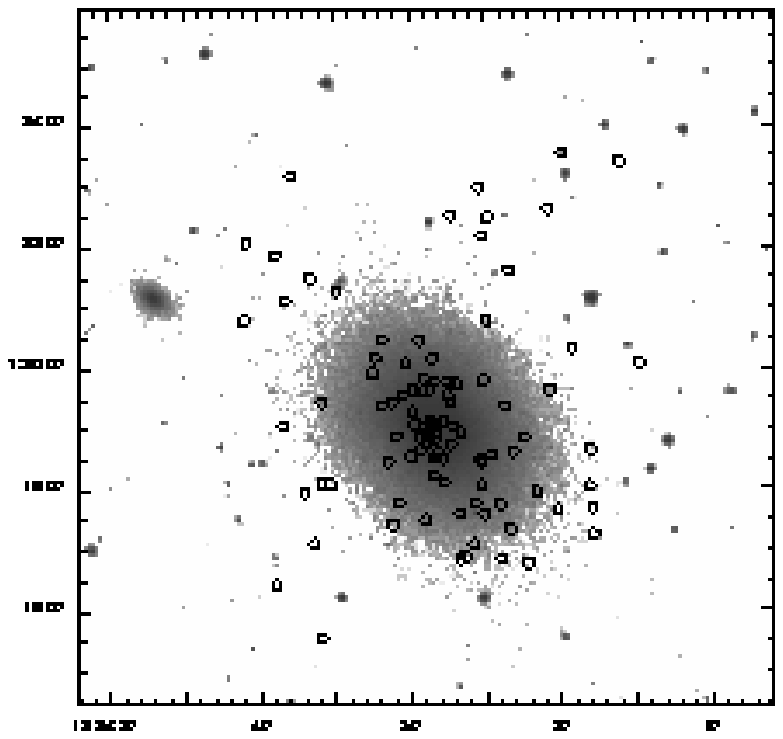}{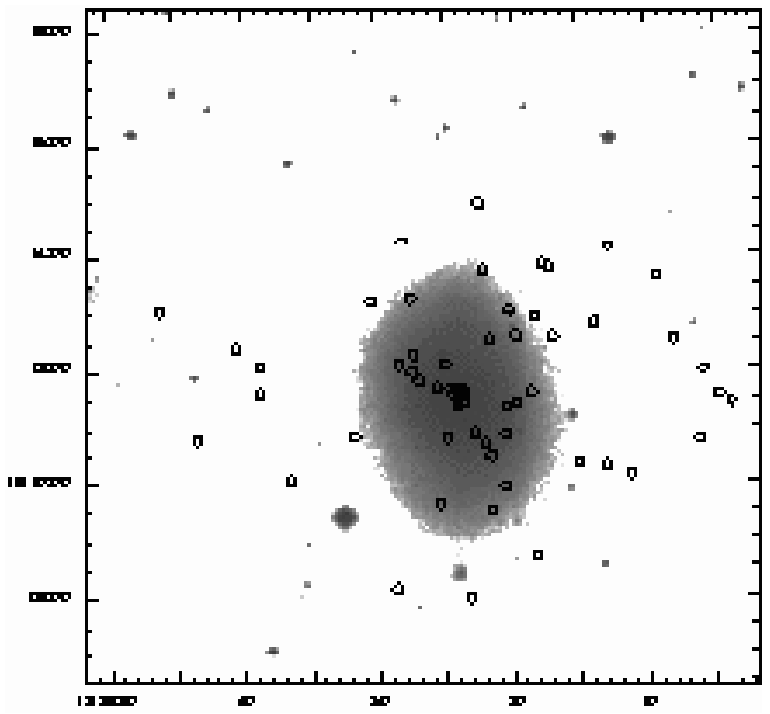}
\caption{
Logarithmic gray scale DSS optical image of (left) NGC~4365 and (right)
NGC~4382. The circles indicate the positions of the X-ray sources listed in
Tables~\ref{tab:src_n4365} and \ref{tab:src_n4382}.
The relative surface brightness range covered by the gray scale is narrower
than in Figure~\protect\ref{fig:adaptive}.
\label{fig:DSS}}
\end{figure}

In Figure~\ref{fig:DSS}, we show the Digital Sky Survey (DSS) optical images
centered on NGC~4365 and NGC~4382 using a logarithmic gray scale. The overlaid
circles indicate the positions of the X-ray point sources listed in
Tables~\ref{tab:src_n4365} and \ref{tab:src_n4382}. Only a small fraction of
the weaker point sources are not evident in Figure~\ref{fig:adaptive}.
In NGC~4365, the diffuse emission qualitatively has the same ellipticity
and position angle (PA) as the optical emission.
On the other hand, the X-ray emission in NGC~4382 appears rounder than
the optical emission.

\section{Resolved Sources} \label{sec:sources}

\subsection{Detections} \label{sec:src_detect}

The discrete X-ray source populations on the ACIS S3 images of the two galaxies
were determined using using the wavelet detection algorithm
({\sc ciao wavdetect}\footnotemark[5] program),
and were confirmed with a local cell
detection method and by visual inspection. The wavelet source detection
threshold was set at $10^{-6}$, which implies that $\la$1 false source
(due to a statistical fluctuation in the background) would be detected
in the entire S3 image.
We further restricted the sources by requiring 
sufficient source counts to determine the source flux at the $\ge 3 \sigma$
level. Over most of the image, the minimum detectable flux was about
$2.7 \times 10^{-4}$ cnt s$^{-1}$ in the 0.3--10 keV band for both galaxies.
The detection limit is slightly higher at large distances
where the Point-Spread-Function (PSF) is larger, and near the center of the
galaxy where confusion with other discrete sources or diffuse gaseous emission
may affect the sensitivity.
For all source analyses, we used a local background with an area three times
that of each source's region
from {\sc wavdetect}.
There were a few cases of nearby sources where the source regions or
background regions were altered slightly to avoid overlaps,
without affecting the net count rates.

Tables~\ref{tab:src_n4365} and \ref{tab:src_n4382} list the discrete sources
detected by this technique in NGC~4365 and NGC~4382, respectively.
In each case, the sources are ordered with increasing projected distance
$d$ from the center of the galaxy.
Columns 1-7 provide the source number, the IAU name, the source
position (J2000),
the projected distance $d$ from the center of each galaxy,
the count rate with its $1 \sigma$ error, and the SNR for the count rate.
The fluxes were corrected for exposure and the instrument PSF.
Since we did not
detect a distinct source which was unambiguously associated with an AGN at the
center of either NGC~4365 or NGC~4382
(see \S~\ref{sec:src_id}),
we adopted the central optical/IR positions of
R.A.\ =\ 12$^{\rm h}$24$^{\rm m}$28\fs26 and
Dec.\ =\ +7\arcdeg19\arcmin3\farcs8 for NGC~4365, and
R.A.\ =\ 12$^{\rm h}$25$^{\rm m}$24\fs04 and
Dec.\ =\ +18\arcdeg11\arcmin25\farcs9 for NGC~4382
from the 2MASS survey \citep{CSV+2001}. As noted in \S~\ref{sec:obs}, the
overall absolute errors are probably $\la$1\arcsec\ for NGC~4365 and
$\sim$0\farcs5 for NGC~4382 near the centers of the fields, with larger errors
further out for both observations.

Our detection limit for sources should result in $\la$1 false source (due to a
statistical fluctuation) in the entire S3 field of view for each galaxy.
However, some of the
detected sources may be unrelated foreground or (more likely) background 
objects. Based on the source counts in \citet{BHS+2000} and \citet{MCB+2000},
we would expect $\approx$12 such unrelated sources in each observation.
These should be spread out fairly uniformly over each S3 image
(Figure~\ref{fig:adaptive}), except for the reduced sensitivity at the
outer edges of the field due to reduced exposure and increased PSF.
Thus, the unrelated sources should mainly be found at larger distances
from the optical centers of the galaxies (the latter entries of
Tables~\ref{tab:src_n4365} \& \ref{tab:src_n4382}), while the sources
associated with the galaxies should be concentrated to their centers.

\subsection{Identifications} \label{sec:src_id}

In addition to the identifications mentioned in \S~\ref{sec:obs}, we visually
examined the first and second generation DSS images for position matches.
Both Src.~98 (USNOA2 0900-07143676) of NGC~4365, and Src.~38 (J1225+182) of
NGC~4382 were visible. In addition, Srcs.~88 and 97 of NGC~4365 and Srcs.~48,
53, and 57 of NGC~4382 had possible optical counterparts. None of these five
sources corresponded to objects listed in NED or SIMBAD.

In both galaxies, the area enclosed by Src.~1 is coincident with our adopted
optical/IR positions of the galaxy center. In NGC~4365, the centroid of
Src.~1 is off by an amount which is consistent with our positional
uncertainty,
while in NGC~4382, the centroid is off by $\ga 3$ times our positional
uncertainty.
It is unclear whether either of these sources is an AGN in the
galaxy center or a LMXB projected near the center.
We also examined hard-band (2--10 keV) images of both galaxies in which
AGNs might be more obvious, particularly as the diffuse emission from
the galaxies is soft.
We did not see any new sources near the center, nor were either of the
centermost sources particularly hard.
Thus, we have no unambiguous evidence for AGNs in these galaxies.
Conservatively, we adopt the luminosity of the centermost
source in each galaxy ($1.5 \times 10^{38}$ ergs s$^{-1}$ in NGC~4365 and
$9.0 \times 10^{37}$ ergs s$^{-1}$ in NGC~4382) as the upper limit to the
central AGN luminosity.

We examined NGC~4382 for a source associated with SN1960r. Src.~27 is the
closest detected source; however, it is $\sim$20\arcsec from the \citet{P1993}
position. We placed a 2.0\arcsec radius circular aperture, similar to that of
Src.~27 plus 0.5\arcsec to account for our positional uncertainty, and a
4.0\arcsec radius circular aperture to determine the local background.
No net counts were detected and we can place a $3 \sigma$
upper limit of $6.3 \times 10^{37}$ ergs s$^{-1}$.

For NGC~4365, A. Kundu kindly provided a list of 325 globular cluster
positions from data presented in \citet{KW2001}.
The WFPC2 field of view overlaps 37
X-ray point sources from Table~\ref{tab:src_n4365}.
After solving for a small
offset, 18 of the 37 sources have a globular cluster within 1\farcs02.
The expected number of associations from random positions is $\sim 0.6$ based
on the density of the globular clusters.
This number is consistent with \citet{KMZ+2003}, who adopted a less strict
source detection criteria.
Recently \citet{LBB+2003} published a short list of globular clusters with
spectroscopy. In addition to globular clusters identifications from
\citet{KW2001}, \citet{LBB+2003} globular cluster \#1 agrees with the position
of our Src.~59.
Sources with candidate globular clusters are noted in
Table~\ref{tab:src_n4365}.
There were no available lists of globular clusters for NGC~4382.

\subsection{X-ray Luminosities and Luminosity Functions} \label{sec:src_lum}

When converting the source count rates into unabsorbed X-ray (0.3--10 keV)
luminosities, we assumed that each source was at the distance of the target
galaxy. We then used the best-fit {\it Chandra} X-ray spectrum
of the inner (one effective radius for NGC~4365, two effective radii for
NGC~4382) resolved sources (Tables~\ref{tab:spectra_n4365} and
\ref{tab:spectra_n4382}, row 3). The conversion factors were
$4.08 \times 10^{41}$ ergs cnt$^{-1}$ for NGC~4365 and
$3.93 \times 10^{41}$ ergs cnt$^{-1}$ for NGC~4382.
Column~8 of Tables~\ref{tab:src_n4365} and \ref{tab:src_n4382} list the
X-ray luminosities in units of $10^{37}$ ergs s$^{-1}$, and range roughly from
$1.1 \times 10^{38}$ to $2.3 \times 10^{39}$ ergs s$^{-1}$ for NGC~4365 and 
$1.1 \times 10^{38}$ to $6.2 \times 10^{39}$ ergs s$^{-1}$ for NGC~4382.

\begin{figure}
\plottwo{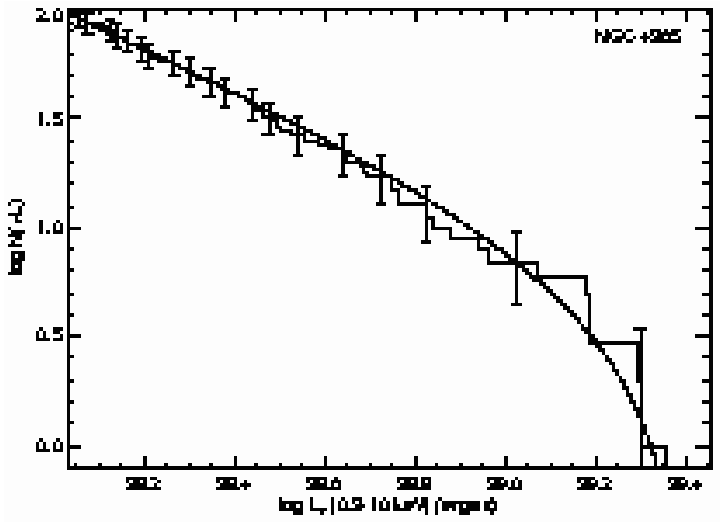}{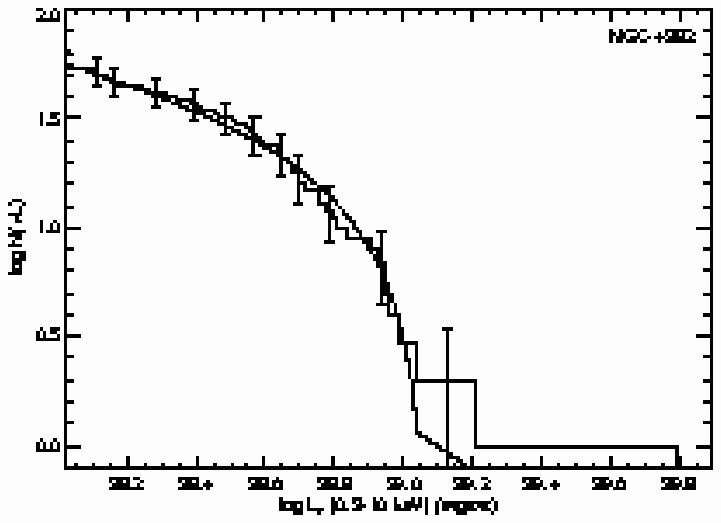}
\caption{
Histogram of the observed cumulative luminosity function of all
resolved sources within the {\it Chandra} S3 fields of (left) NGC~4365 and
(right) NGC~4382. The continuous curves are the sums of the respective
best-fit LMXB luminosity functions (eq. \ref{eq:lfc}) and the expected
background source counts.
Poisson error bars are displayed for every fifth interval.
As this is a cumulative distribution, the errors are correlated.
\label{fig:lf}}
\end{figure}

In Figure~\ref{fig:lf}, we display the cumulative luminosity functions of all
resolved sources in the S3 field. Each
cumulative luminosity function should be the sum of the point source (LMXB)
population of the target galaxy and the foreground/background population. We
fit the luminosity function using the same techniques as we have used
previously
(\citealt{SIB2000,SIB2001}; \citealt*{BSI2001}; \citealt*{ISB2002});
a single power-law, broken
power-law, and a cutoff power-law were all used to model the LMXB population.
The background source population was modeled as discussed in the
previous references.
For NGC~4365, a single power-law fit was acceptable (at $\sim$54\%
confidence level).
However,
the single power-law fit for NGC~4382 was rejected at the greater than 94\%
confidence level.
Attempts at broken power-law fits were more successful
[$\Delta \chi^2 = -6.2$ and $-19.0$, with two less degrees of freedom (dof)
compared to the single power-law];
however, the power-law exponents at high luminosities were steep ($\sim$9), so
that these models effectively had a cutoff above the break luminosity.
As a result, we fit the luminosity functions with a cutoff power-law 
model
($\Delta \chi^2 = -7.3$ and $-19.3$, with one less dof
compared
to the single power-law):
\begin{equation} \label{eq:lfc}
  \frac{ d N }{ d L_{37} } = \left\{
  \begin{array}{ll}
    N_o ( \frac{L_X}{L_c} )^{-\alpha} & L_X \le L_c \\
                                         &                \\
    0                                    & L_X >  L_c  \\
 \end{array}
 \right. \, ,
\end{equation}
where $L_X$ is the X-ray luminosity, $L_{37} \equiv L_X / ( 10^{37}$ erg
s$^{-1})$, and $L_c$ is the cutoff luminosity.
The best fits
were determined by the maximum likelihood method, and the errors
(90\% confidence interval) were determined by Monte Carlo techniques.
For NGC~4365,
$N_o = 2.28^{+4.31}_{-0.59} \times 10^{-2}$,
$\alpha = 1.91^{+0.23}_{-0.30}$, and
$L_c = 226^{+57}_{-90} \times 10^{37}$ ergs s$^{-1}$;
while for NGC~4382,
$N_o = 0.19^{+0.14}_{-0.06}$,
$\alpha = 0.93^{+0.38}_{-0.51}$, and 
$L_c = 110^{+21}_{-23} \times 10^{37}$ ergs s$^{-1}$.

At the faint end of the luminosity function, incompleteness may play a role.
Since the incompleteness is a complicated function of source counts, spectra,
background, the point spread function, and the detection band,
we used simulations to understand the incompleteness.
With
{\sc marx}\footnote{See \url{http://space.mit.edu/CXC/MARX/}.}
simulations, we found that incompleteness
can lead to a depressed luminosity function for luminosities $\sim30\%$ higher
than the minimum detectable luminosity in these galaxies.
In our cases, that corresponds to $\sim
1.4 \times 10^{38}$ ergs s$^{-1}$ (14 counts) for both galaxies. This is in
rough agreement with other reported simulations \citep{KF2003}.
The largest effect of incompleteness for our fits would be in their power-law
slopes.
When we fit the luminosity function at a minimum of $2 \times 10^{38}$ ergs
s$^{-1}$, we find agreement within the errors reported above.

The X-ray luminosity functions of LMXBs have been determined with {\it Chandra}
in a number of other early-type galaxies.
In most cases, the observations are limited to $L_X \ga 10^{38}$ ergs s$^{-1}$.
The best-fit slope $\alpha$ of NGC~4365 is steeper than the low luminosity
slope $\alpha_l$ in
NGC~1553 \citep{BSI2001},
NGC~4649 \citep{RSI2003},
and
NGC~4697 \citep{SIB2001},
although the values for NGC~4365, NGC~4649, and NGC~4697 agree within the 90\%
confidence intervals.
The slope in NGC~4365 is also flatter than the high luminosity slope $\alpha_h$
in these three galaxies, and the single power-law
slope in NGC~1316 \citep{KF2003}.
The 90\% confidence intervals of
NGC~4382 $\alpha$ overlap with $\alpha_l$ in NGC~1553, NGC~4649, and NGC~4697.

The cutoff luminosities of NGC~4365 and NGC~4382
($\approx$ 0.9-3.2 $\times 10^{39}$ ergs s$^{-1}$) are much higher than the
break or cutoff luminosities of NGC~1291, NGC~1553, NGC~4649, and NGC~4697
($\approx$ 2-6 $\times 10^{38}$ ergs s$^{-1}$)
\citep{SIB2001,BSI2001,ISB2002,RSI2003}.
The differences probably are too large to be explained by distance errors.
When we fit the luminosity
functions in NGC~4365 and NGC~4382 with broken power-laws, the break values
were somewhat lower than the cutoff luminosities but consistent within the
errors.
Moreover, the break luminosities were significantly larger than the
values found in NGC~1291, NGC~1553, and NGC~4697.
A very high break luminosity or single power-law fit was found previously
for NGC~1316 \citep{KF2003};
this galaxy had a single power-law fit, but a break at $>5\times 10^{38}$ ergs
s$^{-1}$ could not be excluded.

In \citet{SIB2000}, we argued that a break luminosity similar to the
Eddington luminosity of a $1.4 M_\sun$ spherically accreting neutron star
indicated that the LMXBs above the break contain accreting black holes.
The large luminosities of the cutoff or broken power-laws for NGC~1316,
NGC~4365, and NGC~4382 seem less consistent with this interpretation, barring
a large error in the distance.

In the luminosity function of NGC~4382, the best-fit clearly misses the
brightest source, Src.~38. Since Src.~38 is a bright background AGN, this is
not indicative of a poor fit.

\subsection{Hardness Ratios} \label{sec:src_colors}

\begin{figure}
\plottwo{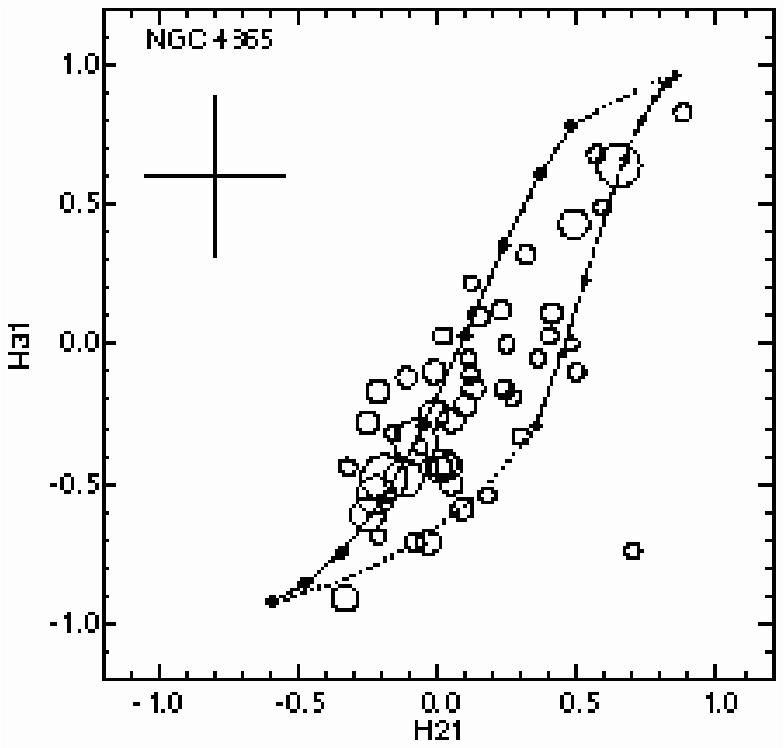}{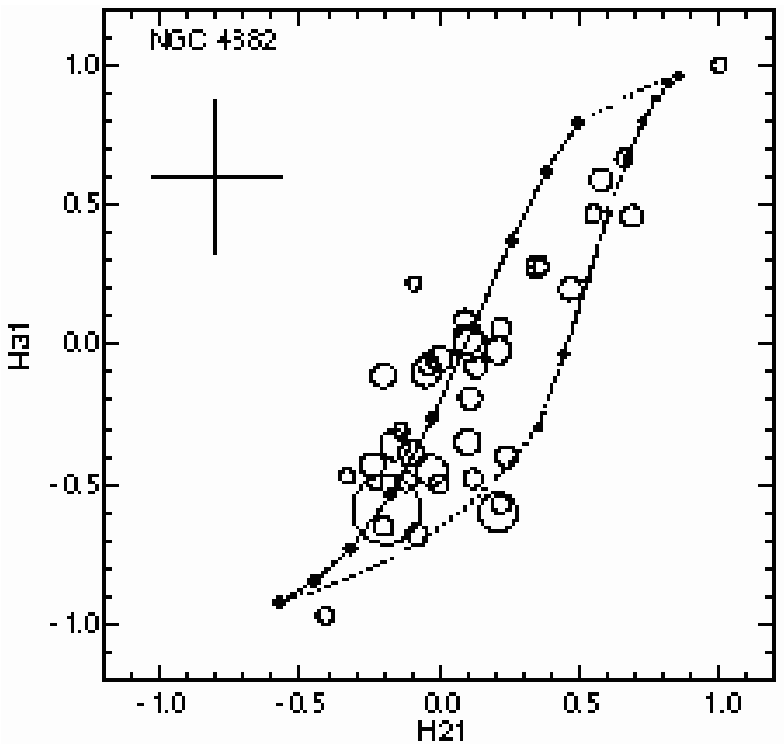}
\caption{
Hardness ratios for the sources of (left) NGC~4365 and (right) NGC~4382 with
at least 20 net counts. Here, $H21 \equiv ( M - S ) / ( M + S )$ and
$H31 \equiv ( H - S ) / ( H + S ) $, where $S$, $M$, and $H$ are the counts in
the soft (0.3--1 keV), medium (1--2 keV), and hard (2--10 keV) bands,
respectively. The area of each circle is proportional to the observed
number of net counts. The solid curve and large diamonds show the hardness
ratios for power-law spectral models with Galactic absorption; the small
diamonds show the ratios for $N_H = 5\times 10^{21}$ cm$^{-2}$; the diamonds
indicate values of the power-law photon number index of $\Gamma = 0$
(upper right) to 3.2 (lower left) in increments of 0.4. The dashed curve
encloses the area with power-law photon indices between 0 and 3.2 and
absorption columns between Galactic and $5\times 10^{21}$ cm$^{-2}$.
The $1 \sigma$ error bars at the upper left illustrate the median of the
uncertainties.
Due to the QE degradation, an effective absorption of $\sim6\times 10^{20}$
cm$^{-2}$ was added to these spectral models.
\label{fig:colors}}
\end{figure}

We determined X-ray hardness ratios for the sources, using the same techniques
and definitions we used previously \citep{SIB2000,SIB2001,BSI2001,ISB2002}.
Hardness ratios or X-ray colors are useful for crudely characterizing the
spectral properties of sources, and can be applied to sources that are too
faint for detailed spectral analysis. We define two hardness ratios as H21
$\equiv ( M - S ) / ( M + S )$  and H31 $\equiv ( H - S ) / ( H + S ) $, where
$S$, $M$, and $H$ are the total counts in the soft (0.3--1 keV), medium
(1--2 keV), and hard (2--10 keV) bands, respectively.
Since the hardness ratios measure observed counts, Galactic absorption, as
well as QE degradation, must be taken into account when interpreting
the ratios.
The hardness ratios and their $1 \sigma$ errors are
listed in columns 9 and 10 of Tables~\ref{tab:src_n4365} and
\ref{tab:src_n4382}. Figure~\ref{fig:colors} plots H31 vs.\ H21 for the 48 and 
39 sources of NGC~4365 and NGC~4382, respectively, with at least 20 net counts.
The hardness ratio for the sum of those sources are (H21,H31) $ = (0.04,-0.25)$
for NGC~4365 and $(0.02,-0.25)$ for NGC~4382.
Sources with $\sim$40 net counts had errors similar to the median of the
uncertainties, $\sim0.26$. The errors scale roughly with the inverse square
root of the net counts.
 
As was also seen in the bulge of NGC~1291, NGC~1553, and NGC~4697
\citep{SIB2000,SIB2001,BSI2001,ISB2002}, most of the sources lie along a broad
diagonal swath extending roughly from (H21,H31) $\approx (-0.3,-0.7)$ to
(0.4,0.5).
In Figure~\ref{fig:colors}, the solid line corresponds to Galactic absorption,
QE degradation, and power-law spectra with photon indices of $\Gamma =  0$ to
3.2.
Most of the sources (37/48 and 26/39) are displaced to the right of the solid
curve. Were the solid curve a good fit to the data, one would expect to find
roughly equal numbers of sources on each side of the curve.
Assuming that this is the case,
the probability of seeing $\le$ 9/48 and 13/39 sources to the
left of the line is 0.01\% and 2.66\% for NGC~4365 and NGC~4382,
respectively. The spectra seem to be more complex than a single power-law with
varying indices.

Between NGC~4697 and the bulge of NGC~1291, there were four sources with
(H21,H31) $\approx (-1,-1)$. These supersoft sources have, essentially, no
emission above 1 keV. Based on scaling to the 3 sources in NGC~4697, one 
might have expected $\sim$5--6 supersoft sources in NGC~4365 and $\sim$3
supersoft sources in NGC~4382.
However, no supersoft sources are observed in either galaxy.

In Figure~\ref{fig:colors}, there are three NGC~4365 sources (Srcs.~36, 69,
and 92) and four NGC~4365 sources with very hard spectra (hardness ratios
[H21,H31] $>$ $[0.5,0.5]$). These may be unrelated, strongly absorbed AGNs,
similar to the sources which produce the hard component of the
X-ray background, and which appear strongly at the faint fluxes in the
deep {\it Chandra} observations of blank fields
\citep{BHS+2000,MCB+2000,GRT+2001}.
There also are two sources with hardnesses ratios of around $(-0.4,-1)$;
these are Src.~38 of NGC~4365 and Src.~10 of NGC~4382; although the hardness
ratio of the latter is highly uncertain.
Studies of other galaxies \citep{SIB2001} and deep blank sky images
\citep[e.g.,][]{GRT+2001} suggest that many of sources, similarly lacking hard
emission, may be unrelated background sources. However, such sources,
especially when at smaller radii, may be part of the host galaxy.
Although the identification of Src.~38 of
NGC~4382 with a $z=1.19$ quasar marks it as a background object, it occupies a
typical part of the figure ([H21,H31] $\approx [-0.2,-0.6]$).

\subsection{Variability} \label{sec:src_var}

\begin{figure}
\plotone{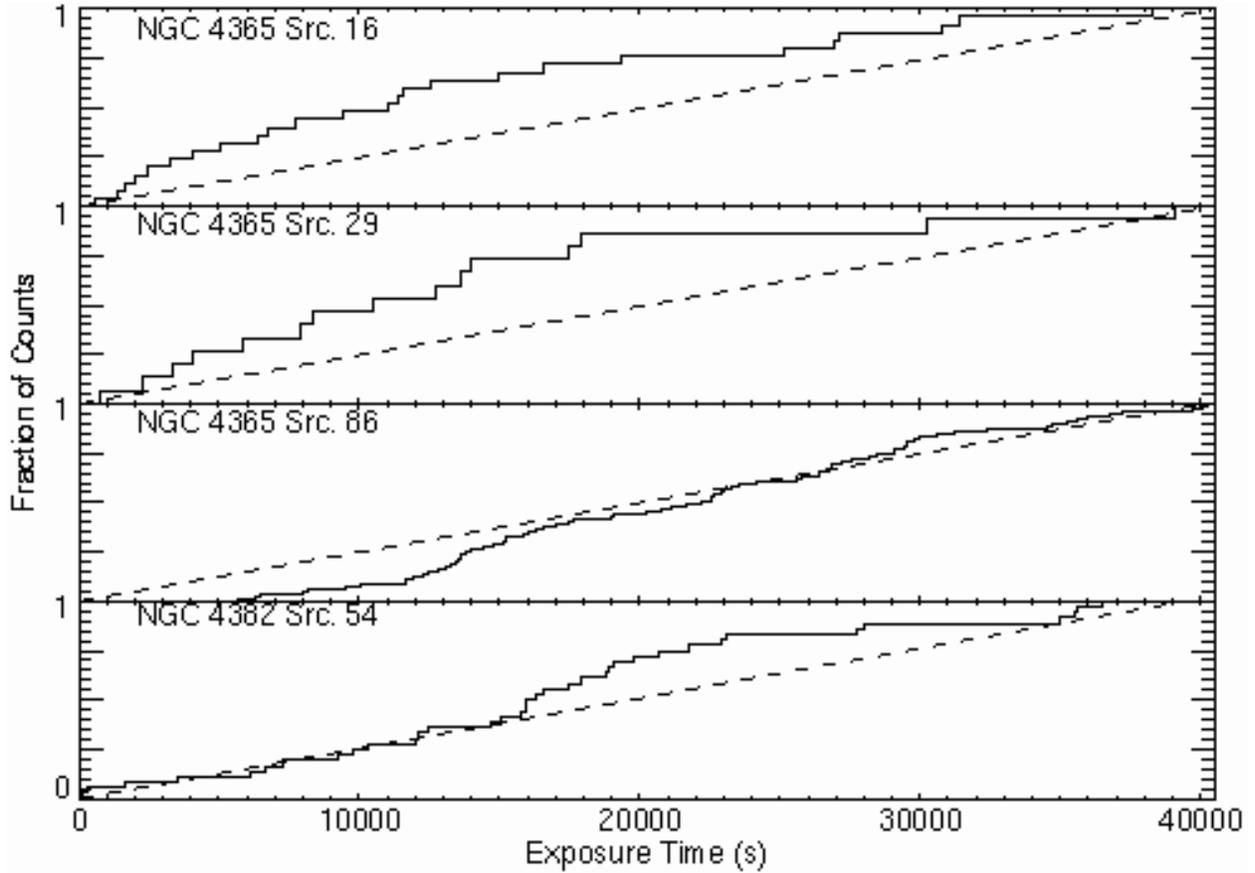}
\caption{
Solid histogram gives accumulated fraction of events for sources as a function
of exposure time. The dashed line is the predicted distribution under
the hypothesis that the source plus background rate is constant. The three
NGC~4365 sources shown are those with variability detected at the $> 99$\%
confidence level. NGC~4382 Src. 54 had variability detected at the $> 96.8$\%
confidence level.
\label{fig:both_variability}}
\end{figure}

\begin{figure}
\plotone{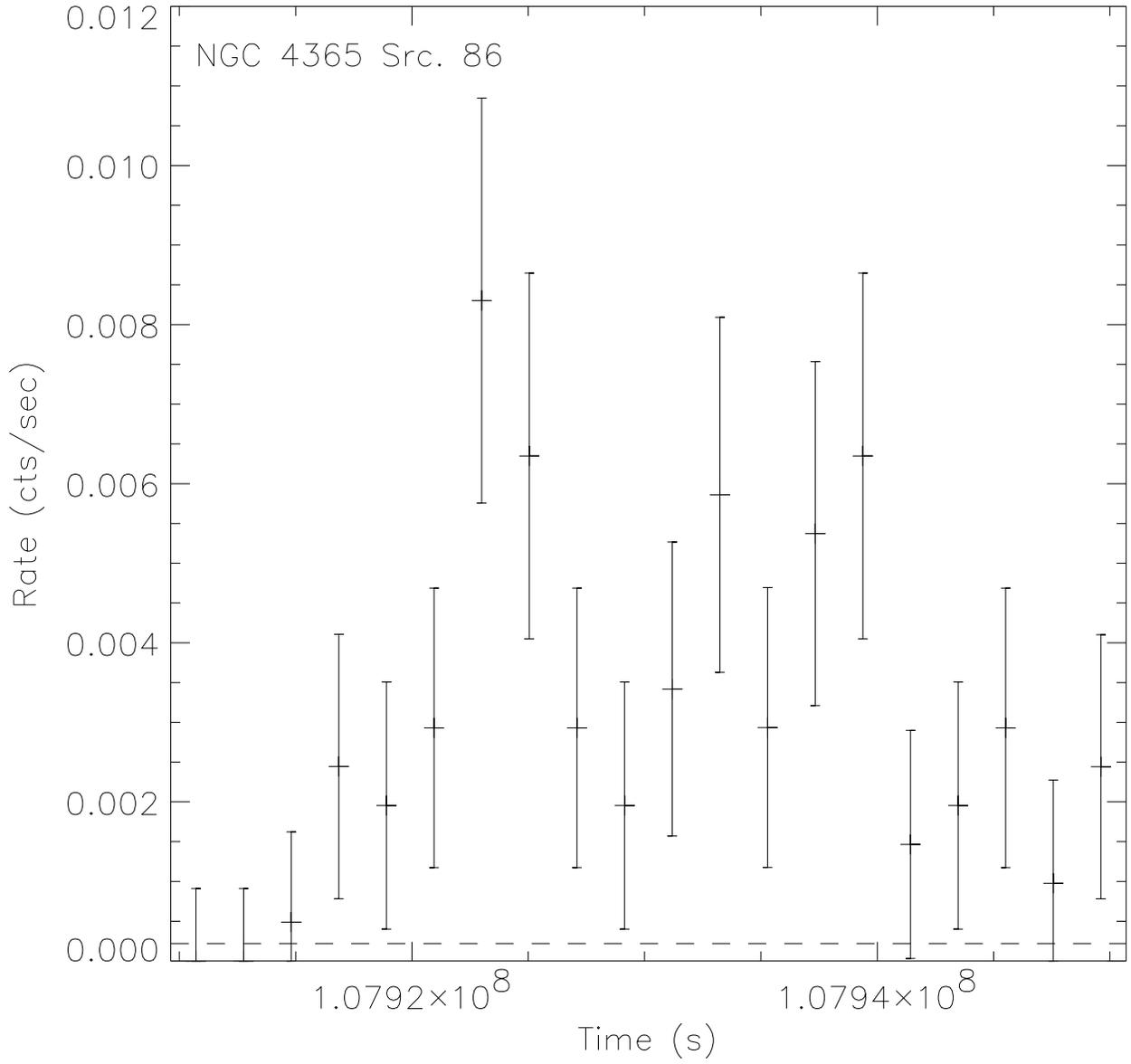}
\caption{
The X-ray light curve of Src.~86 in NGC~4365.
The points with $1 \sigma$ error bars are the source count rate, uncorrected
for background, accumulated in 2021 s time bins.
The horizontal dashed line is the background count rate scaled to the
source region size.
\label{fig:src86}}
\end{figure}

We searched for variability in the X-ray emission of the resolved sources over
the duration of the {\it Chandra} observation using the Kolmogoroff-Smirnov
(KS) test \cite[see][]{SIB2001}. In most cases, the tests were inconclusive.
For NGC~4365, six sources (Srcs.~16, 21, 29, 31, 48, and 86) had a
$> 95$\% probability that they were variable. For NGC~4382, only Src.~54 had a
$> 95$\% probability that it was variable. 
Figure~\ref{fig:both_variability} displays the total count histograms of
the three sources in NGC~4365 (Srcs.~16, 29, and 86) that had a $> 99$\%
probability that they were variable and NGC~4382 Src. 54 which had a $97$\%
probability that it was variable.
NGC~4365 Src.~16 seems to be gradually fading. Similarly, NGC~4365 Src.~29
also seems to fade gradually, but has only 15 net counts.
The X-ray light curve for NGC~4365 Src.~86 is shown in Figure~\ref{fig:src86}.
This source appears to have turned on after the start of the observation,
and undergone a large outburst after about 1/3 of NGC~4365's exposure.
It then decreased, and probably underwent at least one more outburst
before gradually fading as the observation was ending. The light curve
of NGC~4382 Src.~54 suggests that it underwent an outburst after about 3/8 of
NGC~4382's exposure, and that the outburst gradually faded.
Variable sources are noted in Tables~\ref{tab:src_n4365} and
\ref{tab:src_n4382}.

\subsection{Spatial Distribution} \label{sec:src_dist}

As in \citet{SIB2001}, we have performed two comparisons of the spatial
distribution of the X-ray sources with that of the optical light; we have fit
the position angle (PA) distribution and the radial distribution.
For both galaxies, we limited the analysis to all sources within
3$\arcmin$ of the galaxy center to ensure that the circular area lies
completely on the chip. 
In this area, we expect $\sim$5 background sources in each exposure.
We have adopted the Third
Reference Catalogue of Bright Galaxies (RC3) values for the optical
photometry's effective radii, position angle, and ellipticity \citep{VVC+1992}.
At an effective radius of $49\farcs8$, NGC~4365 has PA = 40$\degr$ and
$e=0.28$; while at an effective radius of $54\farcs6$, NGC~4382 has
$e=0.22$. There is no PA listed for NGC~4382 in RC3. \citet{F1997} found that
NGC~4382 has an isophotal twist, with its PA ranging from 60$\degr$ at its
center to 0$\degr$ at large radii. We adopted that latter PA for NGC~4382.
These elliptical isophotes contain one-half of the optical light.

\begin{figure}
\plottwo{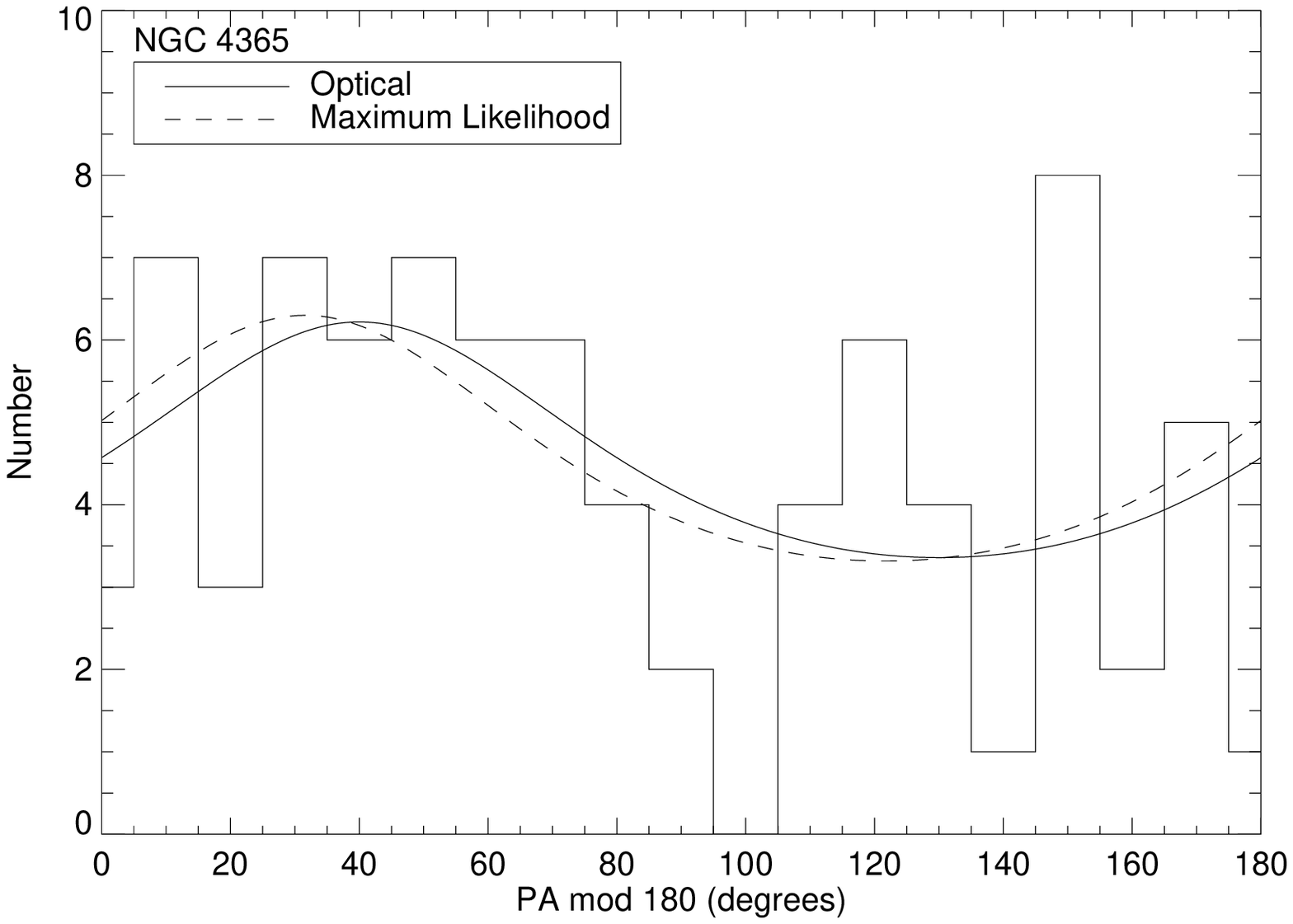}{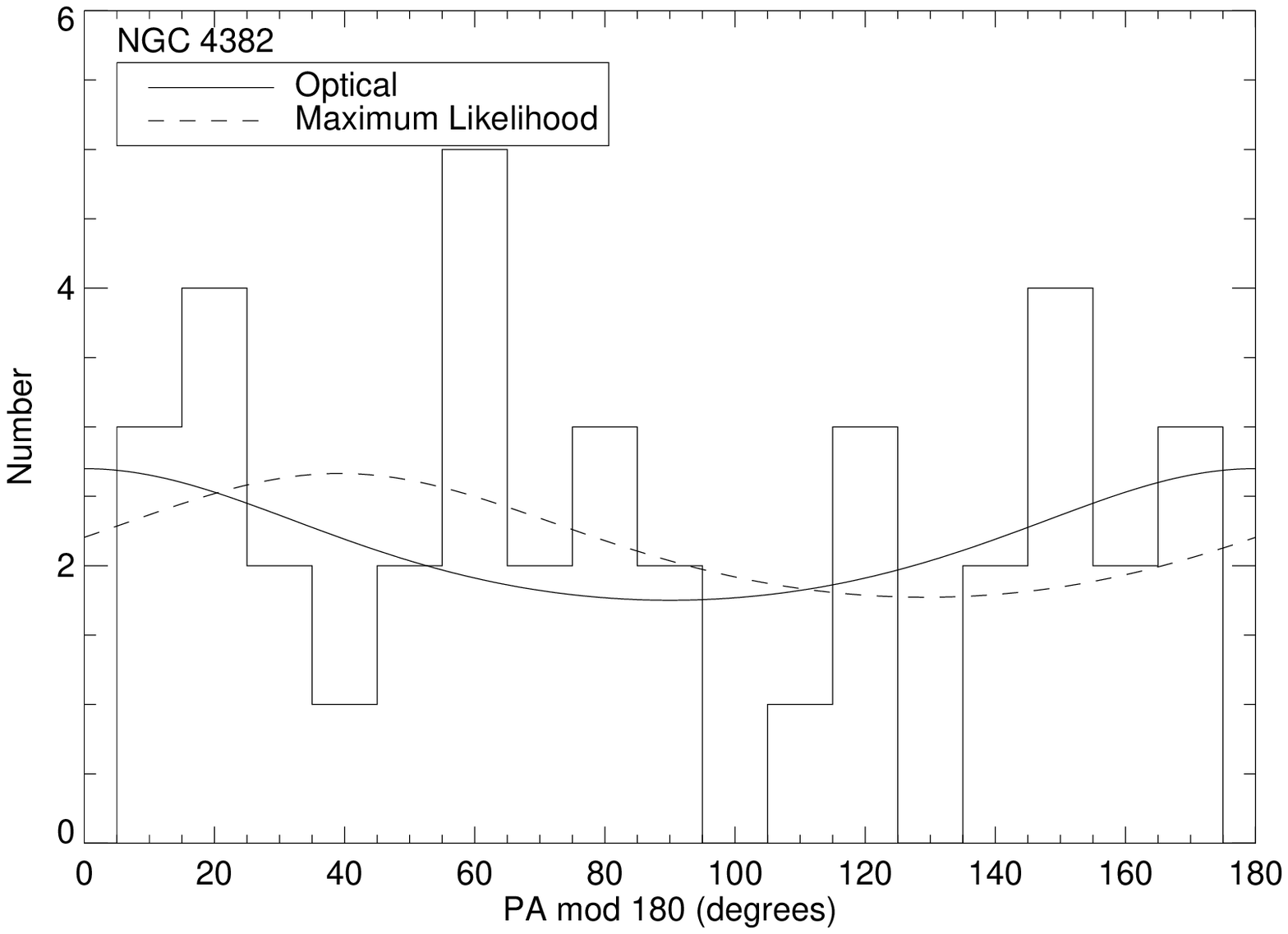}
\caption{
Solid histogram gives the distribution of the position angles of the X-ray
sources within 3$\arcmin$ of the center of (left) NGC~4365 and (right) NGC~4382
in 10$\degr$ bins, as a function of the PA (modulo 180\degr). PA is measured
from north to east. The curves are the expected number of background sources
plus the predicted distribution based on optical photometry and the best-fit
elliptical distribution determined by a maximum likelihood fit to the observed
values.
\label{fig:pa_src}}
\end{figure}

In Figure~\ref{fig:pa_src}, we show the observed distributions
of PAs
(modulo 180\degr, measuring PA from north to east).
In addition to the observed
distribution and the expected optical distribution, we have fit a distribution
representing the number of expected background sources plus the number of
sources expected from a set of sources with constant projected density on
elliptical isophotes.
For NGC~4365 and NGC~4382, $\Delta \chi^2$ between the optical fits and the
maximum likelihood fits indicate that the optical fits are acceptable at the
83\% and 54\% confidence level.
The best-fit distribution of NGC~4365 gives PA = 32\degr\ 
and $e=0.29$, while the best-fit distribution for NGC~4382 has
PA = 39\degr\ and $e=0.21$. The PA value of NGC 4382 lies within the range of
angles \citet{F1997} reports.

\begin{figure}
\plottwo{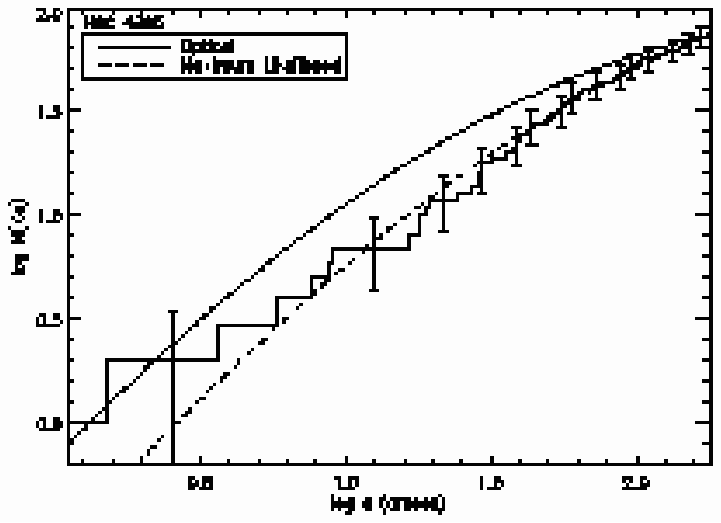}{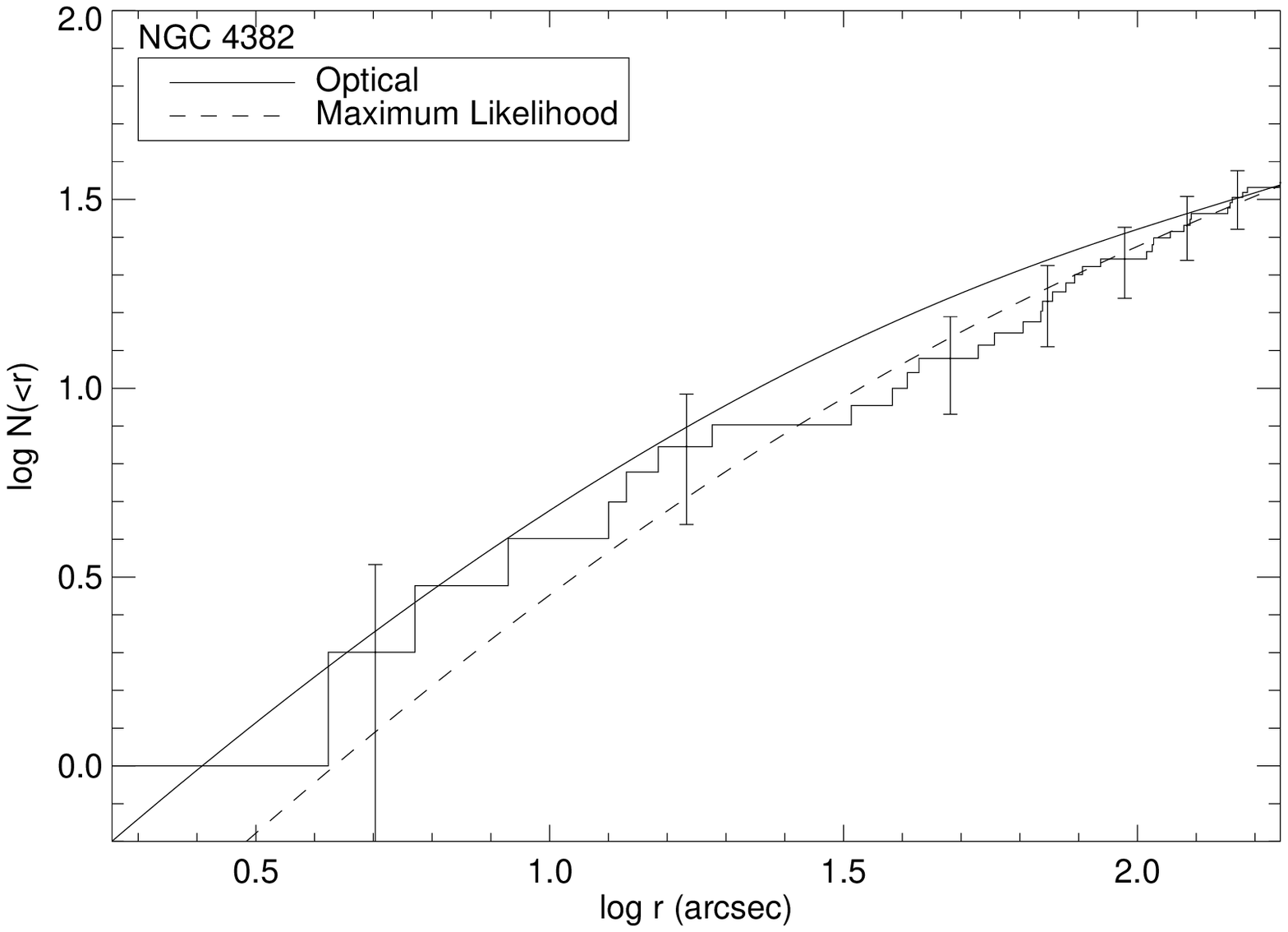}
\caption{
Solid histogram gives the accumulated number of X-ray sources within a
$3\arcmin$ elliptical isophote of the center of (left) NGC~4365 and (right)
NGC~4382 as a function of semi-major axis.
The curves are the expected number of background sources plus the predicted
distribution based on optical photometry and the best-fit distribution
determined by a maximum likelihood fit to the observed values.
Poisson error bars are displayed for every fifth interval.
As this is a cumulative distribution, the $1 \sigma$ error bars are
correlated.
\label{fig:radial_src}}
\end{figure}

Adopting the optically determined values for ellipticity and PA, we show the
accumulated source number as a function of semi-major axis, $a$, in 
Figure~\ref{fig:radial_src}.
The use of elliptical isophotes reduces the number of background sources
modeled in each galaxy by $\sim1$.
We modeled the galactic LMXB distribution using a de Vaucouleurs profile,
and included a uniform distribution of background sources at the level
expected from deep {\it Chandra} observations.
If we fix the effective semi-major axis of the LMXB distribution at the value
determined by the optical light distribution in NGC~4365,
$a_{\rm eff}=58\farcs5$,
the KS test indicates that the optical distribution is rejected at the
$>99$\% confidence level.
The best-fit de Vaucouleurs model has an effective semi-major axis of
136\arcsec.
Similarly,
the KS test for NGC~4382 rejects the optical distribution at the
85\% confidence level.
The best-fit de Vaucouleurs profile for the LMXBs has an effective radius
of 129\arcsec.
Both fits indicate that there are more X-ray sources at larger radii than
expected from the optical light distribution.
Radial fits from circular isophotes yield similar results.
Since NGC~4382 is an S0 galaxy, this disagreement may reflect the
contribution of X-ray sources associated with the disk of the galaxy.
However, this would not explain the difference between the LMXB
distribution and the optical light distribution in NGC~4365. 

We note that a significant fraction of the LMXBs in NGC~4365
are associated with globular clusters
(\S~\ref{sec:src_id}), as is generally true of elliptical galaxies
(\citealt*{ALM2001}; \citealt{SIB2000}).
In elliptical galaxies (including NGC~4365), the globular cluster
population is more broadly distributed than that of the optical light
and field stars
\citep[e.g.,][]{H1991}.
This might help to explain the broader distribution of the LMXBs.
Figure~\ref{fig:radial_src} suggests that there may be two populations
of LMXBs, one with a small effective radius perhaps consistent with
the optical distribution, and one with a larger effective radius.
It is possible that these two distributions represent the LMXBs formed in
globular clusters and those formed from field binary star systems.
However, we note that in NGC~4697, the LMXBs followed the optical light
distribution \citep{SIB2001}.  More elliptical galaxies need to be observed
to resolve this question.

To test whether the LMXBs in these galaxies could discriminate between such
models, we added either a second de Vaucouleurs model or
an exponential model to the optical light distribution.
For either additional component, there were only minimal improvements to
the fits
($\Delta \chi^2 \gtrsim -0.6$ for NGC~4365 and $\gtrsim -0.8$ for NGC~4382,
with one less dof compared to the single profile), indicating
that the current data cannot statistically discriminate between single de
Vaucouleurs distributions with large effective radii or multiple component
distributions.

\section{Spectral Analysis} \label{sec:spectral_analysis}

We extracted spectra of the sources and diffuse emission in both galaxies.
We restricted the spectral analysis to the 0.7 -- 9 keV range.
The lower limit was taken to avoid calibration uncertainties, while
there are few non-background counts beyond 9 keV.
We also found a possible spectral artifact around 1.6--1.9 keV, which
has been noticed by others
\footnote{See
\url{http://asc.harvard.edu/cal/Links/Acis/acis/Cal\_projects/index.html}.}.
We believe this artifact
is minor enough that it should not effect continuum models
(bremsstrahlung or power-law);
however, we chose to excise the 1.6--1.9 keV band when we fit line models
(mekal).
All of the spectra were grouped to have at
least 25 counts per spectral bin prior to background correction to enable our
use of $\chi^2$ statistics.
The use of minimum counts per spectral bin and restricted energy ranges can
result in bins being excluded in the allowed energy range.

The results of the spectral fits are summarized in
Tables~\ref{tab:spectra_n4365} and \ref{tab:spectra_n4382}.
Spectra were extracted for the resolved point sources (`Sources'),
the unresolved diffuse emission excluding the point sources (`Unresolved'),
and for the total emission (sources and unresolved emission).
The total emission spectra were studied for comparison to instruments
with poorer spatial resolution that cannot resolve the point sources.
The third column gives the geometric region for the spectrum;
`Field' implies the entire S3 chip.
The value of the absorbing column density ($N_H$) applied to all components
of the model emission spectrum is given in column 4.
In this and other columns, values in parentheses are fixed (not allowed
to vary).
The fixed value of $N_H$ is the Galactic value from \citet{DL1990}.
As in \citet{SIB2001}, a two component model was necessary to fit the total
spectrum data.
A hard component, modeled by thermal bremsstrahlung (`bremss') or by
a power-law (`power') spectrum, was used to model the resolved and
unresolved LMXBs.
An additional soft line emitting component (`mekal') was used for
gaseous diffuse emission.
We used the {\sc xspec acisabs} model to correct for the QE degradation.
For the hard component,
columns 5--7 give the spectral model,
the temperature $T_h$ (for bremsstrahlung)
or photon number spectral index $\Gamma$, and
the unabsorbed flux of the hard component, $F^h_X$ (0.3--10 keV).
Similarly, columns 8--10 give the temperature $T_s$,
overall heavy element abundance relative to solar, and
flux for the soft mekal component.
For the unresolved emission, the spectra exclude regions around each of the
resolved sources.
The last two columns give the number of net counts in each spectrum,
and $\chi^2$ per degree of freedom (dof) for the best-fit model.
All errors reported in the spectral analysis are 90\% confidence level errors.
Parentheses are used to indicate a frozen parameter and brackets are used when
an error is unconstrained.

The background spectra for the resolved sources were determined locally,
using the same nearby regions as discussed in \S~\ref{sec:sources}.
For the spectra of the resolved diffuse emission and the total spectrum
(sources and unresolved emission), the background was taken from the
deep blank sky backgrounds compiled by Maxim Markevitch\footnotemark[3].

The spectra of several spatial regions were analyzed.
In NGC~4365, the spectra of sources, unresolved emission, and 
total emission were derived from an inner region, which was determined by
the elliptical optical isophote containing one-half of the optical light.
We will refer to this region as within ``one effective radius''.
The semimajor axis of this isophote is defined as $a_{\rm eff}$.
In NGC~4365, this isophote has $a_{\rm eff} = 58\farcs5$,
a semi-minor axis of $42\farcs5$, and a PA of 40\degr\ 
\citep{VVC+1992}.
In NGC~4382, the corresponding isophote did not have sufficient counts from
resolved sources for spectral analysis.
Therefore, we chose the innermost region of NGC~4382 to be twice as large
(2 $a_{\rm eff}$), which gave an elliptical region with a semi-major axis
of 123\arcsec, a semi-minor axis of 95\farcs5, and a PA of 0\degr\ 
\citep{VVC+1992,F1997}.
This isophote contains 69\% of the optical light.
We fit the sources (resolved emission), unresolved
emission, and total emission in these innermost regions.
We also examined the spectra of the sources for the entire field.
In order to search for changing unresolved
emission with radius, we chose to analyze the elliptical annulus between one
and three effective radii for both galaxies, corresponding to $\sim 28.8$\% of
the optical light.
For both galaxies, we used the same PA as the innermost region.
In NGC~4365, there were enough counts from resolved sources between one and
three effective radii to also analyze that spectrum.
Finally, we analyzed the total
emission spectra in the three effective radii region.
Since the different spectra (source, unresolved, and total) were
binned separately, all of the counts for a point source were assigned to a
region based the source was near the edge of the region,
and the source spectra used a local background,
while the unresolved and total spectra used a blank-sky background,
the counts of the source spectrum plus the unresolved spectrum in a region
will not exactly equal the counts of the total spectrum.

\subsection{X-ray Spectra of Resolved Sources} \label{sec:spectra_vs_res}

\subsubsection{NGC~4365}

For NGC~4365, we extracted the spectrum of all resolved sources
within one effective radius.
The observed spectrum is shown on the left side of Figure~\ref{fig:src_spec}.
As has been found in other early-type galaxies \citep{SIB2001},
the combined spectra of the sources was reasonably well-fit by either
a thermal bremsstrahlung model with $kT_h = 6.08$ keV
(Table~\ref{tab:spectra_n4365}, row 1)
or power-law model with a photon number spectral index of $\Gamma = 1.67$
(row 3).
The power-law model gave a slightly better fit, so we adopted
this model as our best-fit.
The fits were not improved significantly when the absorbing column was
allowed to vary (rows 2 \& 4), so we fixed the hydrogen column at the
Galactic value, $N_H = 1.63 \times 10^{20}$ cm$^{-2}$ \citep{DL1990}.
The adopted best-fit power-law model with Galactic absorption and
residuals to the fit are shown on the left side of Figure~\ref{fig:src_spec}.
Although the hardness ratios indicate that each source is probably fit by a
more complex model than a power-law and Galactic absorption, the summation of
counts from objects with different emission properties can be well fit with a
single power-law and Galactic absorption.
Extending the spectrum down to 0.3 keV to see how the restricted
energy range used in fitting affects the absorbing column does result in a
somewhat higher absorption value; however, we believe the calibration at low
energies is still too unreliable to confidently claim an excess absorption.

We also searched for any radial variation in the spectra of the sources.
We extracted the spectrum from an elliptical annulus between
one and three effective radii.
We first fit this region using the source best-fit model from the
inner one effective radius region, which provided an
acceptable fit (row 5).
When we allowed the power-law index to vary (row 6),
we found a better
fit ($\Delta \chi^2 = -2.9$ with one less dof); however, the two
values of $\Gamma$ agree their 90\% confidence limits.
Freeing the absorption also yielded a better fit,
($\Delta \chi^2 = -1.9$ with one fewer dof); however, Galactic
absorption was allowed at the 90\% confidence limit.

We also attempted to fit the collective spectrum of all of the resolved
sources on the S3 chip, which had 2660 net counts.
Again, freeing the power-law index from the best-fit value for the
inner effective radii produced a better fit; however, the power-law index is
still consistent with the best-fit 90\% confidence limit
(rows 8 \& 9).
Freeing the absorbing column did not produce a better fit (rows 9 \& 10).
We find no clear trend between radius and spectral fit.
This is consistent with results from a study of the sources in 15 galaxies
by \citet{IAB2003}.

\subsubsection{NGC~4382}

For NGC~4382, the lower number of sources and counts required using the inner
two effective radii to attain 921 net counts in the spectrum.
For this region, the adopted best-fit model had a power-law index of 1.52
with the Galactic absorption column of $2.45 \times 10^{20}$ cm$^{-2}$
(Table~\ref{tab:spectra_n4382}, row 3). Neither using a bremsstrahlung model
(row 1) nor freeing the absorbing column (rows 2 and 4) gave a better fit.
Again, we remark that although hardness ratios indicate that each source is
probably fit by a more complex model than a power-law and Galactic absorption,
the summation of counts from objects with different emission properties can be
well fit with a single power-law and Galactic absorption.
The best-fit indicated no excess absorption when we extended the spectrum down
to 0.3 keV.

Since there were not enough resolved source counts between one and three
effective radii, we determined the spectrum of the entire field (which of
course includes the inner two effective radius region discussed above).
There were no large statistically-significant differences in the
fits for the entire field and for the inner region (rows 5--7).
Within the 90\% confidence level, the sources could be fit by a power-law
fit with indices between 1.45 and 1.65 (inclusive of the inner
two effective radius fit) and an absorbing column of $\la 11 \times 10^{20}$
cm$^{-2}$.
Note that an important contribution to the spectrum in the outer parts of
the observation of NGC~4382 is Src.~38, which is a known background AGN.
However, this AGN has X-ray hardness ratios which indicate that it
doesn't have a particularly hard or absorbed spectrum
(Table~\ref{tab:src_n4382}).
The X-ray hardness ratios for this source are similar to the average values
for the LMXBs in NGC~4382.
Since we can rarely identify an AGN with a particular source, AGN always
contribute to spectra describing the resolved sources. Removing Src.~38 from
the spectral analysis would hinder comparisons with other galaxies' resolved
sources.

\subsection{X-ray Spectra of Unresolved Emission} \label{sec:spectra_vs_unres}

\subsubsection{NGC~4365}

\begin{figure}
\epsfig{file=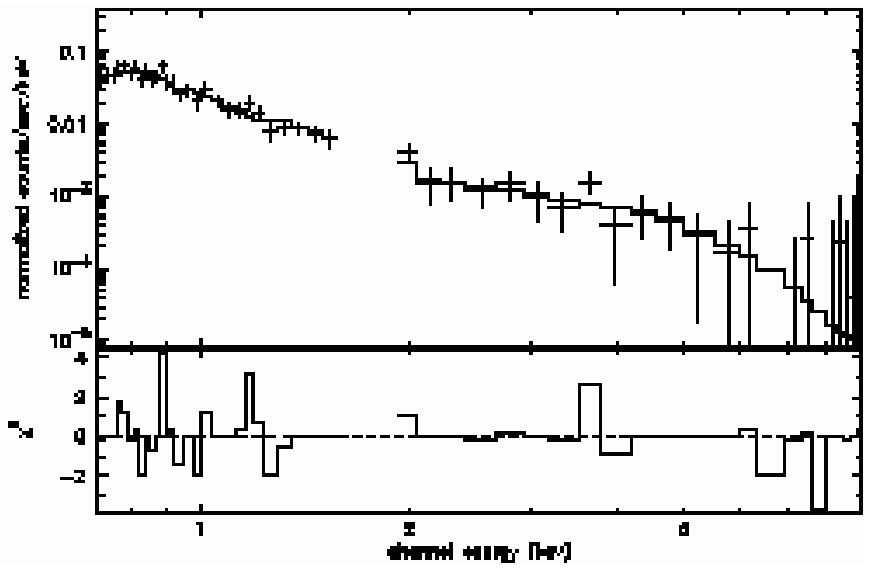,width=0.5\textwidth,clip=}
\epsfig{file=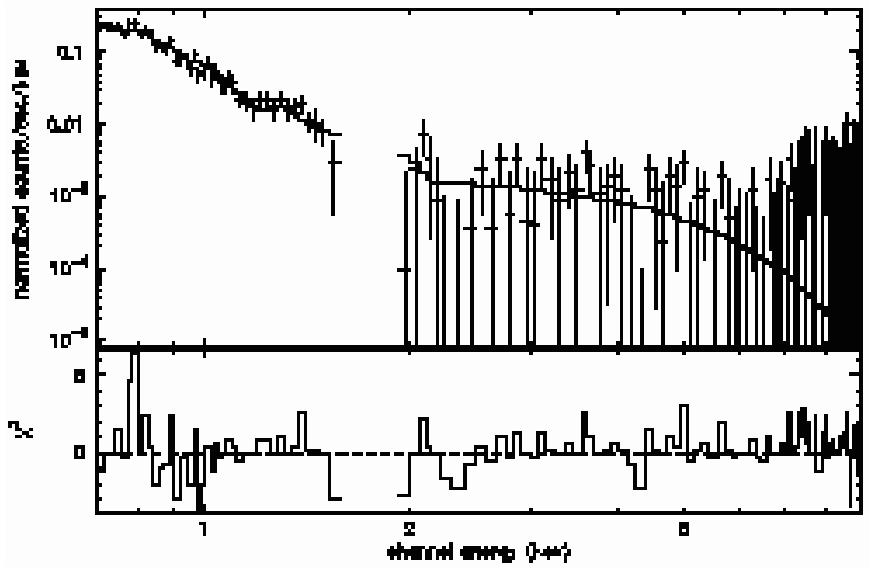,width=0.5\textwidth,clip=}
\caption{
In the upper panels, the cumulative X-ray spectra of the unresolved
emission in the inner effective radius of NGC~4365 (left) and inner two
effective radii of NGC~4382 (right) with $1 \sigma$ error bars
are overlayed by the solid
histograms of the best-fit model spectra (Table~\ref{tab:spectra_n4365} row 13
and \ref{tab:spectra_n4382} row 11). Lower panels display the contribution to
$\chi^2$ with the sign indicating the sign of the residual.
\label{fig:unres_spec}}
\end{figure}

Again, we began by exploring the spectrum of the inner effective radius, which
had 935 net counts.
This spectrum is shown on the left side of Figure~\ref{fig:unres_spec}.
First, we attempted to model the unresolved emission with a
soft mekal component representing the emission by diffuse interstellar gas
(Table~\ref{tab:spectra_n4365}, row 11).
In general, none of the spectral fits were improved by allowing the
absorbing column to vary (e.g., row 12), so we will discuss only fits with
a fixed Galactic column.
Values from both mekal-only fits are consistent with previous ROSAT
measurements from \citet{DW1996}.
The fits to the unresolved emission using only a soft mekal model showed
large residuals at high energies.
Of course, the unresolved emission includes unresolved point sources as
well as diffuse gas, and the unresolved point sources will have a hard spectrum
if they are like the resolved sources.
Thus, we added a hard component to the fit with the same spectral shape as
the best-fit model for the source.
This produced a dramatic reduction in $\chi^2$ by 37 for one less dof,
clearly indicating values from single temperature fits, like in \citet{DW1996},
may be suspect.
Allowing the power-law index to vary did not improve the fit
significantly.
Therefore, we adopted the model with a power-law index of 1.67, a mekal
temperature of $0.56^{+0.05}_{-0.08}$ keV, and an abundance
in solar units of $0.35$ (row 13) as our best-fit model for
unresolved emission (Figure~\ref{fig:unres_spec}).
Although the line emission strength from heavy elements is well constrained,
minimal emission from hydrogen leads to poorly constrained ($>0.08$)
abundances.
These best-fit values are consistent with some of the soft temperatures and
abundances found from two-temperature fits to ASCA observations \citep{BF1998}.

We also examined whether the unresolved emission's spectrum changed with
radius. The unresolved spectrum between one and three effective radii 
had 1155 net counts. We first fit this spectrum with the best-fit model for
unresolved emission (row 14).
Then, we allowed the soft component model to vary in the fit.
This produced a much better fit ($\Delta \chi^2 = -37$) with a statistically
significant higher temperature (row 15).
This suggests that there is a positive radial gradient in the temperature
of the gas.
If there is a significant gradient, this would also affect the spectra
in the inner parts of the galaxy.
We would be observing both cooler gas located near the center of the
galaxy, and hotter gas at larger radii seen in projection against the
center.
Thus, we tried fitting the
emission from the inner effective radius with two soft components and
a hard component;
that model did not improve on the single soft+hard model
spectral fit we report in Table~\ref{tab:spectra_n4365}.
A temperature gradient
may exist in the diffuse gas; however, deeper observations of NGC~4365 are
necessary to do a more detailed deprojection.

In addition to a possible temperature gradient in the diffuse emission, the
spectral fits of the sources and unresolved emission point to a possible
change in the unresolved sources with respect to radius. In the resolved
sources, the model flux for sources in the inner effective radius and sources
between one and three effective radii are about the same. The correction for
increased contribution by background sources based on the luminosity function
we fit earlier suggests that the flux from resolved sources between one and
three effective radii should be $\sim 0.9$ times the flux from resolved sources
in the inner effective radius. The luminosity function of background
sources in a
particular field can differ from an average field either by total number of
sources or by having a brighter than expected background source in the field.
Source 69 is a very hard and bright source that might be such an unusually
bright AGN. This source accounts for $\sim1/4$ of the hard counts in the one
to three effective radii resolved source region. If we were to exclude this
source, the flux ratio between the two regions would be $\sim0.65$.
One would expect that if the unresolved sources distribution is the same as
the resolved source distribution, we should find a flux ratio between
$\sim 0.6$ and $1.0$ for the unresolved sources. Instead, the best-fit
unresolved emission between one and three effective radii is a purely soft
component fit. At best, we can limit the ratio to be $\lesssim 0.15$. This
suggests that the unresolved sources radial distribution may decline more
rapidly than the resolved source distribution. Poisson errors in the 2--9 keV
count rate of the one to three effective radii unresolved emission are too
large to statistically confirm this result. Deeper X-ray observations of
NGC~4365 are needed to resolve enough sources to allow the faint source
and bright source radial distribution to be compared.

\subsubsection{NGC~4382}

The spectrum of the unresolved emission of the inner two effective radii
(right side of Figure~\ref{fig:unres_spec}) had 2620 counts.
We initially fit the spectrum with a model containing only a soft mekal
component representing emission from diffuse interstellar gas
(row 8).
The best-fit models with varying absorption gave values below the
Galactic column, but with an upper limit which included the Galactic value
(Table~\ref{tab:spectra_n4382}, row 9).
Thus, we chose to use the Galactic column for all remaining fits of this
galaxy's spectra.
As with NGC~4365, the addition of a hard component representing unresolved
sources led to a much better statistical fit (row 10).
Allowing the parameters of the hard component to vary did not result in
a significantly better fit, and the hard component parameters were very
poorly constrained.
Thus, we fixed the shape of the spectrum of the hard component to the
adopted model for the resolved sources in the galaxy.
This leads to an acceptable fit (row 10); however, its abundance is abnormally
high and poorly constrained due to minimal emission from hydrogen. Therefore,
we have frozen the abundance at 0.5 solar for all remaining soft component
fits in this galaxy. This choice (row 11) is very close to the best fit
($\Delta \chi^2 = 0.3$). This best-fit model has a temperature of
$0.28^{+0.04}_{-0.03}$ and is shown in Figure~\ref{fig:unres_spec}.

We extracted the unresolved emission spectrum from a larger (but overlapping)
elliptical annulus extending from one to three effective radii.
The spectrum in this region was reasonably fit by the same gas temperature
and abundance which provided the best fit within two effective radii (row 12).
When the gas temperature was allowed to vary (row 13), the best-fit
temperature increased; however, that increase was not statistically
significant.
Unlike in NGC~4365, we find no change in the flux ratio of the
resolved and unresolved sources with increasing radius.

\subsection{Total X-ray Spectra} \label{sec:spectra_total}

\subsubsection{NGC~4365}

We first extracted the spectrum of all of the emission within one
effective radius.
The spectrum in this region is consistent with the best-fit spectra for
the resolved and unresolved emission (row 16);
when the spectral parameters of the hard and soft components were allowed
to vary, the fits were not significantly improved (rows 17 and 18) and did not
constrain the abundance.
We also attempted a model with a bremsstrahlung hard component (row 19). This
fit slightly better than the power-law model, was more successful at
constraining the abundance ($<0.26$ solar), but did not constrain the
bremsstrahlung temperature well ($>5.21$ keV). We therefore adopted the sum of 
the best-fit spectra for the resolved and unresolved emission as our best-fit
for the total emission.
The total emission within three effective radii yielded the same fit within the
overlap of the two region's errors for a power-law hard component model. Again,
a bremsstrahlung hard component could better constrain the soft component
abundance at the expense of constraining the hard component temperature.
The abundance is significantly lower than the best-fit unresolved abundance;
however, the statistics can not discriminate between the power-law fit, with
its poorly constrained abundance ($>0.01$ solar), and the bremsstrahlung fit to
within a 90\% confidence level. Therefore, we do not claim any abundance
gradient in the total emission.

In \citet{MKA+1997}, a bremsstrahlung model with a temperature of
$12.0^{+29.3}_{-\phn5.5}$ keV or a power-law model with index
$1.8^{+0.4}_{-0.4}$ was fit to the hard component, while the soft component
was fit by either a mekal or a Raymond-Smith model with temperature $\sim0.5$
keV and poorly constrained abundance ($>0.016$ solar). These results are
consistent with what we have found.

\subsubsection{NGC~4382}

The results for NGC~4382 were similar to those for NGC~4365.
Within two effective radii, the total spectrum was fit well
by the combination of the best-fit hard component from the resolved
sources and the best-fit soft component from the diffuse emission
(Table~\ref{tab:spectra_n4382}, row 14).
Freeing the hard and soft spectral parameters did not improve the
fit significantly (rows 15 and 16).
A bremsstrahlung hard component did not produce a better fit and poorly
constrained the bremsstrahlung temperature $>10.38$ keV. There was no
significant change between the total emission within two effective radii and
the total emission within three effective radii.

Once again, the results were similar to \citet{MKA+1997}. They used the same
hard component model discussed above for NGC~4365, and fit a soft temperature
$\sim0.3$ keV. \citet{MKA+1997} were able to constrain the abundance using
the Raymond-Smith model to 0.029--0.21 solar, but could not constrain the
abundance for the mekal model.

\subsection{Discrete versus Diffuse Emission}
\label{sec:src_vs_dif}

We used both counts (0.3--10.0 keV) and spectral fitting results to
determine the fraction of the X-ray emission which was due to diffuse gas
versus the fraction due to point sources, whether resolved or not. The
unresolved emission is a combination of unresolved point sources and truly
diffuse gaseous emission. We used the resolved source hardness ratios of the
each region to separate unresolved counts into point sources and diffuse gas.
We also compared the hard flux (point sources) to the soft flux (diffuse gas)
in the total emission spectra. Uncertainty in the spectral models used lead to
a much wider range of derived flux ratios than count ratios.
Since the gas is
softer than the sources, absorption affects the diffuse emission more.
Therefore, the unabsorbed flux ratio of point sources to total emission need
not be larger than the count ratio of point sources to total emission, as one
one would expect.

For the inner one effective radius of NGC~4365 the point sources are
responsible for $75\pm7\%$ of the total counts and $\sim70\%$ of the
unabsorbed flux (Table~\ref{tab:spectra_n4365}, rows 16--19). Approximately
$60\%$ of the point source counts are resolved. The unresolved point source
flux overlaps with 90\% confidence level fluxes obtained from the derived
luminosity function extended down to $10^{36}$ ergs s$^{-1}$.
Moving to the region within three effective radii the point sources are
responsible for $75\pm6\%$ of the total counts and $\sim55\%$ of the
unabsorbed flux (rows 20 \& 21).

In the inner two effective radii of NGC~4382 the point sources account
for $55\pm5\%$ of the total counts and $\sim50\%$ of the unabsorbed flux
(Table~\ref{tab:spectra_n4382}, rows 14--17), with $\sim40\%$ of the point
source counts resolved.  Again, the unresolved point source flux overlaps with
90\% confidence level fluxes obtained from the derived luminosity function
extended down to $10^{36}$ ergs s$^{-1}$. In the inner three effective radii,
the point sources account for $56\pm6\%$ of the total counts and $\sim50\%$ of
the unabsorbed flux (rows 18 \& 19).

For both galaxies, it was crucial that a majority of the source were resolved.
For power-law models, the 90\% confidence limit on the power-law index was cut
by a factor of 2--3 sources were resolved. If one uses
bremsstrahlung models, the bremsstrahlung temperature had no upper constraint
unless the sources' spectra could be fit separately.

\section{Spatial Distribution of the Diffuse Gaseous Emission}
\label{sec:dif_dist}

We derived the radial distribution of the diffuse gas, based on the
surface brightness distribution in the soft band (0.3--1 keV).
We excluded regions around each of the resolved sources in determining the
diffuse surface brightness.
Although the diffuse gas generally dominates in the soft band, this emission
still contains a contribution from unresolved sources
(\S~\ref{sec:spectra_vs_unres}).
To remove the effect of the unresolved point sources, we have
assumed that the spectrum of the sum of all resolved sources in the
S3 field is representative of the spectrum of the unresolved sources.
Based on the observed spectra, we also assume that the diffuse gas produces
no significant hard band (2--10 keV) emission.
Thus, we determined the surface brightness profile in the hard band
(removing resolved sources), and scale this by the ratio of soft to hard 
band emission expected based on the best-fit spectrum of the resolved
point sources.
This soft band surface brightness profile for the sources was
subtracted from the observed soft band profile, and we attribute the
remainder to diffuse gas.
We performed this operation in regular circular annuli out to
$\sim 3 \arcmin$, and corrected the profiles for background and exposure.

\begin{figure}
\plottwo{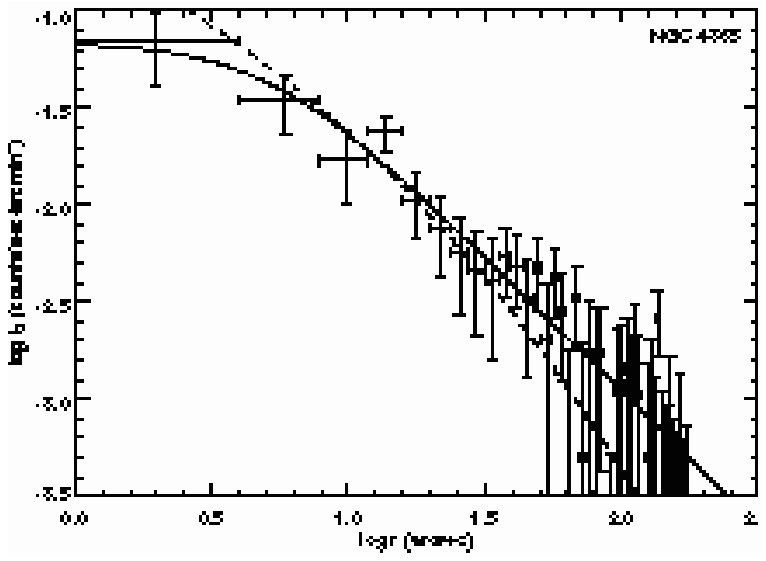}{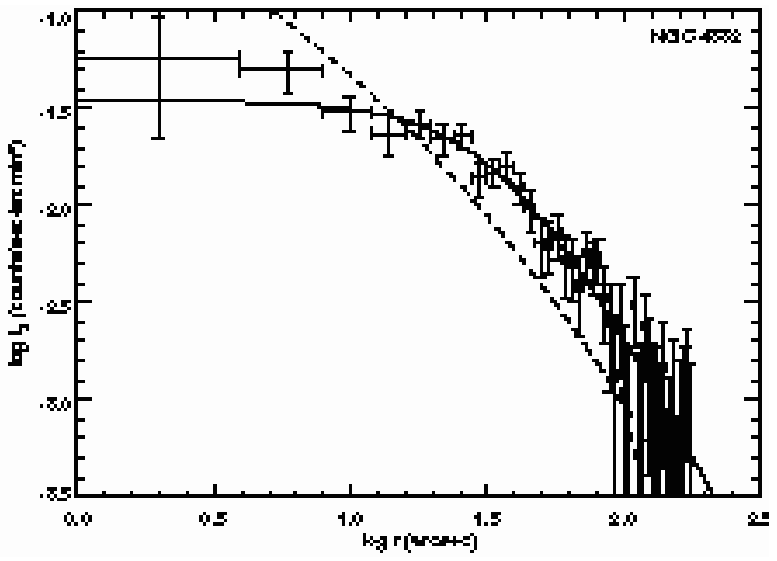}
\caption{
Surface brightness profiles of the soft band (0.3--1 keV) diffuse gaseous
emission of (left) NGC~4365 and (right) NGC~4382 as a function of projected
radius $r$ with $1 \sigma$ error bars.
The dashed curve shows the RC3 de Vaucouleurs profile with an effective radius
determined from the distribution of optical light in the galaxy, but
with the normalization varied to fit the X-ray surface brightness.
The solid curve is the best-fit beta model. All fits were for
$ r < 2 \arcmin$.
\label{fig:diffuse_sb}}
\end{figure}

Figure~\ref{fig:diffuse_sb} displays the observed soft band gaseous surface
brightness profiles as a function of projected radius, $r$.
The dashed lines display the best-fit de Vaucouleurs profiles with effective
radii fixed at the values for the optical light in the galaxies
(49$\farcs$8 for NGC~4365 and 54$\farcs$6 for NGC~4382).
The normalization was varied to achieve the best fit.
NGC~4365 is acceptably fit by such a profile, with
a $\chi^2$ of 29.8 for 29 dof.
NGC~4382 is unacceptably fit by the de Vaucouleurs profile,
and the diffuse gas surface brightness is consistently larger than the optical
profile at $r \gtrsim 16 \arcsec$.
\citet*{BBA1998} found that NGC~4382 could be decomposed into a bulge with
an effective radius of $71.4 \arcsec$ and a disk with a scale length of
$165.5\arcsec$. Adjusting the normalization of the bulge plus disk profile
better fit the X-ray profile; however, it was still an unacceptable
fit ($\chi^2 > 99$ for 29 dof).
Therefore, we attempted a standard beta model profile,
\begin{equation} \label{eq:beta}
  I_X(r) = I_0\left[1+\left(\frac{r}{r_c}\right)^2\right]^{-3\beta+1/2},
\end{equation}
where $r_c$ is the core radius.
The fit for NGC~4365 was much better than the
de Vaucouleurs profile, $\chi^2$ of 17.7 for 27 dof with
$r_c = 5\farcs1^{+8.7}_{-3.6}$ (0.50 kpc) and
$\beta = 0.398^{+0.083}_{-0.045}$ (90\% confidence errors).
This near power-law fit is similar to those for NGC~1291 and NGC~4697
\citep{ISB2002,SIB2001}. NGC~4382 had a similar quality beta model fit with
$r_c = 35\farcs5^{+21.8}_{-13.0}$ (3.2 kpc) and $\beta = 0.599\pm0.265$.
Since $\beta$ is constrained by the outer data points, the much larger core
radius of NGC~4382 is responsible for the larger confidence limit of its
$\beta$ compared to NGC~4365.
The beta models for both galaxies are similar to the profiles of X-ray bright
galaxies within the errors \citep{FJT1985,TFC1986}.
For NGC~4382, its large core radius, rounder X-ray emission than optical
emission, and lower gas temperature at inner radii are roughly consistent with 
predictions by \citet{BM1996} for a rotating interstellar gas distribution;
however, the core radius is about half their predicted value for a galaxy in
Virgo.

\section{Conclusions} \label{sec:conclusion}

The ability of \textit{Chandra} observations to separate X-ray emission into
point sources (LMXBs and background sources) and diffuse emission (unresolved
LMXBs and diffuse gas) plays a crucial part in exploring the properties of the
X-ray faint early-type galaxies NGC~4365 and NGC~4382.
We detected 99 sources in NGC~4365 and 58 sources in NGC~4382, while we
expect that approximately 12 sources in each
galaxy are due to unrelated background sources.
Within one effective radius of
NGC~4365, $\sim 45\%$ of the counts is resolved into sources,
$\sim 30\%$ of the counts is attributed to unresolved LMXBs, and $\sim 25\%$
is attributed to diffuse gas. This distribution is consistent with that of
another X-ray faint elliptical galaxy, NGC~4697 \citep{SIB2001}, after
correcting for different sensitivities.
Within two effective radii of NGC~4382, $\sim 22\%$ of the counts
is resolved into sources, $\sim 33\%$ of the flux is attributed to unresolved
LMXBs, and $\sim 45\%$ is attributed to diffuse gas. The higher diffuse gas
fraction in this lenticular is consistent with NGC~1553 \citep{BSI2001}, also
an X-ray faint lenticular galaxy; however, NGC~4382 resolves less of its flux
into sources than NGC~1553, despite a better sensitivity.

The hardness ratios of the resolved sources tend to be harder (more H21) than
a range of power-law indices with Galactic absorption.
This suggests that the source emission spectra are more complex than single
power-laws.
No supersoft sources were found in either of the two galaxies.
The spectra of the sum of all the sources were best fit
by a power-law model with Galactic absorption.
The power-law of NGC~4382, 1.52, was harder than that of NGC~4365, 1.67;
however, the indices are the same within their 90\% confidence interval and
are consistent with the best-fit value of the sources simultaneously fit in a
survey of 15 galaxies \citep{IAB2003}.

For sources in NGC~4365, 18 sources out of 37 in a WFPC2 field of view had
positions correlated with a globular cluster compared to the $\sim0.6$ expected
from random association.
An additional source's position has also been identified with a globular
cluster.
Three of the sources in NGC~4365 are variable at the $>$99\% confidence
level.
The spatial distributions of the sources are broader than the
optical de Vaucouleurs distribution of each galaxy.

The most surprising results from analysis of the sources are the luminosity
fits. Both galaxies were best fit with cutoff power-laws, where the cutoff
luminosities were between $\approx$ 0.9-3.1 $\times 10^{39}$ ergs s$^{-1}$.
Although broken power-laws were within the 90\% confidence interval, the
break luminosity was similar to the cutoff luminosity. This is in contrast to
the prior break/cutoff luminosities of NGC~1291, NGC~1553, and NGC~4697
($\approx$ 2-6 $\times 10^{38}$ ergs s$^{-1}$), which had been identified
with the Eddington luminosity of a $1.4 \, M_\sun$ spherically accreting
neutron star \citep{SIB2001,BSI2001,ISB2002}.
It seems unlikely that the cutoff luminosities of NGC~4365 and NGC~4382
can be attributed to that physical mechanism.
One other X-ray faint early-type galaxy, NGC~1316, indicates a break luminosity
that may be as high as the cutoff luminosities found in NGC~4365 \& NGC~4382;
however, the galaxies in this paper provide a tighter constraint on a 
cutoff/break luminosity well above the neutron star Eddington luminosity.

In NGC~4365, the temperature of the diffuse gas is $\sim 0.6$ keV and 
its abundance is poorly constrained ($>0.08$ solar) in the inner effective
radius.
There is some evidence for a positive temperature gradient in NGC~4365;
however, deeper observations are necessary to confirm any radial gradients in
temperature or abundance.
NGC~4382 was best fit with a temperature of $\sim 0.28$ keV and a poorly
constrained abundance ($>0.13$ solar).

The diffuse gas surface brightness profiles in both galaxies were fit by
standard beta model profiles.
NGC~4365 had a small ($5\arcsec$, 0.5 kpc) core radius and $\beta \sim 0.40$,
similar to NGC~4697.
NGC~4382 had a larger core radius $\sim 36\arcsec$ (3.2 kpc) and $\beta$
between $\sim 0.33$ and $0.86$.
The larger core radius, combined with the X-ray emission being rounder than
the optical emission, and evidence of a cooler central region, may indicate
rotation of the diffuse gas.

\acknowledgements
We are very grateful to Arunav Kundu for providing us with his
unpublished list of globular clusters in NGC~4365, and for
several very helpful conversations.
We thank Mike Skrutskie for his help and advice in using the 2MASS
catalog.
Support for this work was provided by the National Aeronautics and Space
Administration through $Chandra$ Award
Numbers
GO1-2078X
and
GO1-3100X,
issued by the $Chandra$ X-ray Observatory Center, which is operated by the
Smithsonian Astrophysical Observatory for and on behalf of NASA under
contract NAS8-39073.
This research has made use of the NASA/IPAC Extragalactic Database (NED) which
is operated by the Jet Propulsion Laboratory, California Institute of
Technology, under contract with the National Aeronautics and Space
Administration, the SIMBAD database, operated at CDS, Strasbourg, France, and
the Digitized Sky Surveys, produced at the Space Telescope Science Institute
under U.S. Government grant NAG W-2166.

\bibliography{ms}

\clearpage
\begin{table}[p]
\begin{center}
\caption{Discrete X-ray Sources in NGC~4365 \label{tab:src_n4365}}
\tiny{
\begin{tabular}{lcccrrrrccl}
\tableline
\tableline
Src.&
&
R.A.&
Dec.&
\multicolumn{1}{c}{$d$}&
\multicolumn{1}{c}{Count Rate}&
&
&
&
&
\\
No.&
Name&
(h:m:s)&
($\arcdeg$:$\arcmin$:$\arcsec$)&
\multicolumn{1}{c}{($\arcsec$)}&
\multicolumn{1}{c}{($10^{-4}$ s$^{-1}$)}&
\multicolumn{1}{c}{SNR}&
\multicolumn{1}{c}{$L_X$}&
H21&
H31&
Notes \\
(1)&
(2)&
(3)&
(4)&
\multicolumn{1}{c}{(5)}&
\multicolumn{1}{c}{(6)}&
\multicolumn{1}{c}{(7)}&
\multicolumn{1}{c}{(8)}&
(9)&
(10)&
(11)
\\
\tableline
 1&CXOU J122428.3$+$071904&12 24 28.32& 7 19 04.1&\phn\phn1.0&   37.34$\pm$3.18&   11.75&   152.49&$-0.12^{+0.12}_{-0.11}$&$-0.48^{+0.13}_{-0.11}$&a,b,d\cr
 2&CXOU J122428.1$+$071902&12 24 28.18& 7 19 02.9&\phn\phn1.4&   14.23$\pm$1.98&\phn7.17&\phn58.10&$-0.25^{+0.23}_{-0.21}$&$-0.28^{+0.23}_{-0.20}$&a,b\cr
 3&CXOU J122428.4$+$071906&12 24 28.44& 7 19 06.0&\phn\phn3.5&   48.89$\pm$3.61&   13.53&   199.64&$-0.08^{+0.09}_{-0.09}$&$-0.35^{+0.11}_{-0.10}$&a,b\cr
 4&CXOU J122428.5$+$071905&12 24 28.59& 7 19 05.5&\phn\phn5.2&\phn8.86$\pm$1.56&\phn5.69&\phn36.16&$+0.23^{+0.34}_{-0.41}$&$+0.12^{+0.38}_{-0.42}$&a,b\cr
 5&CXOU J122427.8$+$071907&12 24 27.89& 7 19 07.3&\phn\phn6.5&\phn4.59$\pm$1.16&\phn3.95&\phn18.75&$-0.38^{+0.56}_{-0.38}$&$-0.37^{+0.59}_{-0.39}$&d,f\cr
 6&CXOU J122428.3$+$071857&12 24 28.36& 7 18 57.0&\phn\phn7.0&\phn5.94$\pm$1.35&\phn4.39&\phn24.26&$+0.70^{+0.25}_{-0.77}$&$-0.74^{+1.74}_{-0.26}$&\cr
 7&CXOU J122428.6$+$071908&12 24 28.67& 7 19 08.0&\phn\phn7.4&   13.66$\pm$1.97&\phn6.93&\phn55.76&$-0.21^{+0.24}_{-0.22}$&$-0.17^{+0.23}_{-0.22}$&\cr
 8&CXOU J122428.7$+$071853&12 24 28.70& 7 18 53.7&   \phn12.0&\phn3.37$\pm$1.01&\phn3.34&\phn13.76&$-0.08^{+0.57}_{-0.52}$&$-0.22^{+0.66}_{-0.51}$&d,f\cr
 9&CXOU J122428.9$+$071856&12 24 28.97& 7 18 56.6&   \phn12.8&\phn7.11$\pm$1.43&\phn4.98&\phn29.04&$+0.36^{+0.26}_{-0.34}$&$-0.05^{+0.43}_{-0.41}$&d,f\cr
10&CXOU J122429.1$+$071859&12 24 29.18& 7 18 59.6&   \phn14.3&\phn5.40$\pm$1.28&\phn4.23&\phn22.04&$+0.48^{+0.30}_{-0.47}$&$+0.00^{+0.57}_{-0.57}$&d,f\cr
11&CXOU J122428.4$+$071849&12 24 28.43& 7 18 49.4&   \phn14.6&\phn3.53$\pm$1.04&\phn3.39&\phn14.42&$+0.06^{+0.46}_{-0.48}$&$-0.36^{+0.73}_{-0.45}$&\cr
12&CXOU J122427.1$+$071901&12 24 27.18& 7 19 01.6&   \phn16.2&\phn7.18$\pm$1.45&\phn4.95&\phn29.33&$+0.27^{+0.29}_{-0.35}$&$-0.19^{+0.49}_{-0.41}$&d,f\cr
13&CXOU J122429.3$+$071850&12 24 29.33& 7 18 50.4&   \phn20.8&\phn4.16$\pm$1.14&\phn3.67&\phn17.00&$-0.74^{+0.87}_{-0.22}$&$-0.38^{+0.52}_{-0.35}$&d,f\cr
14&CXOU J122429.7$+$071904&12 24 29.72& 7 19 04.1&   \phn21.7&   16.48$\pm$2.12&\phn7.78&\phn67.31&$+0.05^{+0.17}_{-0.17}$&$-0.27^{+0.22}_{-0.19}$&\cr
15&CXOU J122428.9$+$071844&12 24 28.92& 7 18 44.0&   \phn22.1&\phn6.53$\pm$1.31&\phn5.00&\phn26.66&$+0.18^{+0.28}_{-0.32}$&$-0.54^{+0.61}_{-0.32}$&\cr
16&CXOU J122427.2$+$071844&12 24 27.29& 7 18 44.4&   \phn24.2&\phn5.74$\pm$1.30&\phn4.42&\phn23.45&$-0.06^{+0.34}_{-0.33}$&$-0.37^{+0.46}_{-0.33}$&g\cr
17&CXOU J122426.6$+$071856&12 24 26.61& 7 18 56.3&   \phn25.7&\phn3.22$\pm$0.97&\phn3.31&\phn13.14&$+0.61^{+0.30}_{-0.71}$&$-0.20^{+1.00}_{-0.71}$&d,f\cr
18&CXOU J122427.3$+$071926&12 24 27.31& 7 19 26.1&   \phn26.4&   11.78$\pm$1.80&\phn6.56&\phn48.11&$+0.10^{+0.21}_{-0.22}$&$-0.22^{+0.28}_{-0.24}$&d,f\cr
19&CXOU J122429.8$+$071916&12 24 29.85& 7 19 16.6&   \phn26.9&   14.33$\pm$1.97&\phn7.28&\phn58.51&$-0.03^{+0.17}_{-0.17}$&$-0.71^{+0.25}_{-0.15}$&\cr
20&CXOU J122428.1$+$071836&12 24 28.17& 7 18 36.0&   \phn27.8&\phn4.53$\pm$1.13&\phn4.00&\phn18.50&$+0.06^{+0.42}_{-0.44}$&$-0.03^{+0.48}_{-0.46}$&\cr
21&CXOU J122429.3$+$071839&12 24 29.32& 7 18 39.0&   \phn29.3&\phn3.89$\pm$1.07&\phn3.63&\phn15.88&$+0.08^{+0.54}_{-0.60}$&$+0.04^{+0.56}_{-0.59}$&g\cr
22&CXOU J122427.4$+$071934&12 24 27.49& 7 19 34.4&   \phn32.7&\phn3.93$\pm$1.06&\phn3.72&\phn16.04&$+0.29^{+0.37}_{-0.48}$&$+0.08^{+0.48}_{-0.52}$&\cr
23&CXOU J122428.5$+$071831&12 24 28.50& 7 18 31.1&   \phn32.9&   21.17$\pm$2.38&\phn8.90&\phn86.44&$+0.02^{+0.15}_{-0.15}$&$-0.43^{+0.19}_{-0.15}$&\cr
24&CXOU J122427.8$+$071830&12 24 27.88& 7 18 30.5&   \phn33.8&   13.64$\pm$1.93&\phn7.06&\phn55.69&$-0.01^{+0.21}_{-0.21}$&$-0.10^{+0.24}_{-0.23}$&\cr
25&CXOU J122428.7$+$071938&12 24 28.72& 7 19 38.7&   \phn35.5&\phn6.86$\pm$1.38&\phn4.97&\phn27.99&$+0.25^{+0.27}_{-0.32}$&$+0.00^{+0.36}_{-0.36}$&d,f\cr
26&CXOU J122429.1$+$071937&12 24 29.13& 7 19 37.9&   \phn36.5&\phn2.69$\pm$0.89&\phn3.04&\phn11.00&$+0.38^{+0.40}_{-0.62}$&$+0.06^{+0.61}_{-0.66}$&d,f\cr
27&CXOU J122429.9$+$071832&12 24 29.97& 7 18 32.5&   \phn40.4&\phn3.51$\pm$1.01&\phn3.48&\phn14.32&$+0.33^{+0.37}_{-0.51}$&$-0.16^{+0.67}_{-0.55}$&\cr
28&CXOU J122430.9$+$071852&12 24 30.92& 7 18 52.8&   \phn41.2&\phn2.79$\pm$0.90&\phn3.10&\phn11.40&$-0.14^{+0.60}_{-0.51}$&$+0.11^{+0.46}_{-0.51}$&\cr
29&CXOU J122429.8$+$071938&12 24 29.82& 7 19 38.5&   \phn41.7&\phn2.65$\pm$0.89&\phn3.00&\phn10.83&$+0.29^{+0.42}_{-0.58}$&$+0.05^{+0.57}_{-0.60}$&g\cr
30&CXOU J122428.4$+$071946&12 24 28.42& 7 19 46.9&   \phn43.2&\phn6.67$\pm$1.36&\phn4.91&\phn27.23&$-0.32^{+0.32}_{-0.26}$&$-0.44^{+0.36}_{-0.26}$&\cr
31&CXOU J122427.6$+$071947&12 24 27.63& 7 19 47.1&   \phn44.3&\phn2.66$\pm$0.89&\phn3.00&\phn10.87&$+0.00^{+0.47}_{-0.47}$&$-0.87^{+1.87}_{-0.13}$&g\cr
32&CXOU J122430.4$+$071933&12 24 30.48& 7 19 33.6&   \phn44.5&\phn4.05$\pm$1.08&\phn3.74&\phn16.54&$+0.31^{+0.38}_{-0.51}$&$+0.08^{+0.53}_{-0.58}$&\cr
33&CXOU J122426.8$+$071944&12 24 26.89& 7 19 44.7&   \phn45.7&\phn5.03$\pm$1.51&\phn3.33&\phn20.53&$+0.45^{+0.30}_{-0.45}$&$-0.73^{+1.67}_{-0.27}$&\cr
34&CXOU J122429.1$+$071949&12 24 29.10& 7 19 49.3&   \phn47.2&\phn2.94$\pm$0.93&\phn3.17&\phn11.99&$-0.55^{+0.75}_{-0.34}$&$-0.68^{+0.93}_{-0.28}$&d,f\cr
35&CXOU J122431.2$+$071925&12 24 31.20& 7 19 25.7&   \phn49.0&\phn5.08$\pm$1.20&\phn4.24&\phn20.76&$-0.59^{+0.35}_{-0.21}$&$-0.77^{+0.46}_{-0.17}$&\cr
36&CXOU J122428.4$+$071813&12 24 28.48& 7 18 13.9&   \phn50.0&\phn7.66$\pm$1.48&\phn5.18&\phn31.27&$+0.88^{+0.11}_{-0.92}$&$+0.83^{+0.16}_{-1.06}$&\cr
37&CXOU J122425.3$+$071828&12 24 25.35& 7 18 28.0&   \phn56.2&\phn2.79$\pm$0.92&\phn3.04&\phn11.40&$-0.43^{+0.73}_{-0.41}$&$-0.43^{+0.73}_{-0.41}$&d,f\cr
38&CXOU J122427.7$+$071806&12 24 27.74& 7 18 06.9&   \phn57.4&   18.65$\pm$2.27&\phn8.23&\phn76.14&$-0.33^{+0.15}_{-0.13}$&$-0.91^{+0.26}_{-0.07}$&\cr
39&CXOU J122431.8$+$071923&12 24 31.89& 7 19 23.4&   \phn57.5&\phn3.22$\pm$0.96&\phn3.37&\phn13.15&$+0.48^{+0.40}_{-0.79}$&$+0.37^{+0.48}_{-0.82}$&\cr
40&CXOU J122425.2$+$071826&12 24 25.24& 7 18 26.6&   \phn58.3&\phn3.60$\pm$1.03&\phn3.48&\phn14.69&$-0.02^{+0.38}_{-0.38}$&$-0.83^{+1.42}_{-0.17}$&d,f\cr
41&CXOU J122431.4$+$071827&12 24 31.45& 7 18 27.5&   \phn59.8&\phn5.43$\pm$1.22&\phn4.45&\phn22.19&$-0.16^{+0.35}_{-0.31}$&$-0.32^{+0.40}_{-0.31}$&\cr
42&CXOU J122424.6$+$071833&12 24 24.66& 7 18 33.9&   \phn61.3&   11.86$\pm$1.79&\phn6.63&\phn48.44&$+0.13^{+0.21}_{-0.22}$&$-0.16^{+0.28}_{-0.25}$&\cr
43&CXOU J122425.1$+$071947&12 24 25.16& 7 19 47.9&   \phn63.9&\phn3.95$\pm$1.06&\phn3.74&\phn16.11&$+0.50^{+0.32}_{-0.57}$&$+0.24^{+0.51}_{-0.68}$&\cr
44&CXOU J122428.5$+$072009&12 24 28.56& 7 20 09.5&   \phn65.8&\phn3.22$\pm$1.07&\phn3.02&\phn13.14&$+0.00^{+0.45}_{-0.45}$&$-0.67^{+1.09}_{-0.30}$&d,f\cr
45&CXOU J122430.3$+$072005&12 24 30.30& 7 20 05.1&   \phn68.4&\phn4.21$\pm$1.19&\phn3.55&\phn17.21&$+0.24^{+0.41}_{-0.53}$&$+0.15^{+0.46}_{-0.54}$&d,f\cr
46&CXOU J122423.7$+$071922&12 24 23.73& 7 19 22.7&   \phn69.9&\phn2.62$\pm$0.86&\phn3.05&\phn10.70&$+0.00^{+0.45}_{-0.45}$&$-0.67^{+1.09}_{-0.30}$&\cr
47&CXOU J122425.2$+$071804&12 24 25.29& 7 18 04.6&   \phn73.9&   10.72$\pm$1.68&\phn6.39&\phn43.76&$+0.09^{+0.20}_{-0.21}$&$-0.59^{+0.35}_{-0.22}$&\cr
48&CXOU J122423.1$+$071836&12 24 23.15& 7 18 36.9&   \phn80.6&\phn2.70$\pm$0.89&\phn3.02&\phn11.01&$+0.20^{+0.54}_{-0.70}$&$+0.06^{+0.64}_{-0.69}$&d,f,g\cr
49&CXOU J122432.5$+$071955&12 24 32.50& 7 19 55.2&   \phn81.5&\phn3.17$\pm$0.96&\phn3.29&\phn12.94&$+0.50^{+0.36}_{-0.70}$&$+0.40^{+0.43}_{-0.74}$&\cr
50&CXOU J122429.4$+$072027&12 24 29.42& 7 20 27.1&   \phn85.1&\phn6.29$\pm$1.31&\phn4.79&\phn25.69&$-0.21^{+0.31}_{-0.27}$&$-0.69^{+0.59}_{-0.23}$&d,f\cr
51&CXOU J122430.7$+$071746&12 24 30.75& 7 17 46.5&   \phn85.7&\phn3.65$\pm$1.03&\phn3.55&\phn14.89&$+0.09^{+0.41}_{-0.44}$&$-0.50^{+0.84}_{-0.40}$&\cr
52&CXOU J122425.6$+$071746&12 24 25.69& 7 17 46.0&   \phn86.7&   16.83$\pm$2.10&\phn8.02&\phn68.71&$+0.01^{+0.16}_{-0.16}$&$-0.45^{+0.22}_{-0.18}$&\cr
53&CXOU J122422.4$+$071851&12 24 22.40& 7 18 51.4&   \phn88.0&\phn5.85$\pm$1.26&\phn4.64&\phn23.89&$+0.59^{+0.25}_{-0.45}$&$+0.49^{+0.31}_{-0.51}$&d,f\cr
54&CXOU J122432.2$+$072010&12 24 32.29& 7 20 10.5&   \phn89.7&\phn4.01$\pm$1.05&\phn3.81&\phn16.37&$-0.14^{+0.39}_{-0.35}$&$-0.50^{+0.57}_{-0.32}$&\cr
55&CXOU J122426.7$+$071736&12 24 26.76& 7 17 36.6&   \phn90.0&\phn4.84$\pm$1.14&\phn4.25&\phn19.75&$-0.75^{+0.73}_{-0.21}$&$-0.63^{+0.49}_{-0.24}$&\cr
56&CXOU J122429.0$+$071729&12 24 29.03& 7 17 29.8&   \phn94.7&\phn5.73$\pm$1.24&\phn4.62&\phn23.39&$+0.12^{+0.36}_{-0.39}$&$+0.22^{+0.32}_{-0.38}$&\cr
57&CXOU J122431.8$+$072028&12 24 31.85& 7 20 28.2&   \phn99.9&\phn2.81$\pm$0.88&\phn3.20&\phn11.50&$-0.25^{+0.49}_{-0.39}$&$-1.00^{+2.00}_{-0.00}$&\cr
58&CXOU J122425.1$+$071734&12 24 25.13& 7 17 34.9&      100.4&   10.92$\pm$1.70&\phn6.42&\phn44.58&$+0.41^{+0.20}_{-0.24}$&$+0.11^{+0.29}_{-0.31}$&\cr
					    			    
59&CXOU J122424.0$+$071745&12 24 24.04& 7 17 45.1&      100.7&\phn3.35$\pm$0.96&\phn3.49&\phn13.67&$-0.33^{+0.57}_{-0.40}$&$-0.13^{+0.47}_{-0.42}$&d,f\cr
60&CXOU J122431.1$+$071724&12 24 31.18& 7 17 24.7&      108.2&\phn7.62$\pm$1.42&\phn5.35&\phn31.11&$+0.02^{+0.30}_{-0.30}$&$+0.03^{+0.31}_{-0.31}$&\cr
\tableline
\end{tabular}}
\end{center}
\end{table}

\clearpage

\begin{table}[p]
\begin{center}
\tiny{
\begin{tabular}{lcccrrrrccl}
\multicolumn{11}{c}{TABLE \protect\ref{tab:src_n4365}---$Continued$}\cr
\tableline
\tableline
Src.&
&
R.A.&
Dec.&
\multicolumn{1}{c}{$d$}&
\multicolumn{1}{c}{Count Rate}&
&
&
&
&
\\
No.&
Name&
(h:m:s)&
($\arcdeg$:$\arcmin$:$\arcsec$)&
\multicolumn{1}{c}{($\arcsec$)}&
\multicolumn{1}{c}{($10^{-4}$ s$^{-1}$)}&
\multicolumn{1}{c}{SNR}&
\multicolumn{1}{c}{$L_X$}&
H21&
H31&
Notes \\
(1)&
(2)&
(3)&
(4)&
\multicolumn{1}{c}{(5)}&
\multicolumn{1}{c}{(6)}&
\multicolumn{1}{c}{(7)}&
\multicolumn{1}{c}{(8)}&
(9)&
(10)&
(11)
\\
\tableline
61&CXOU J122424.9$+$072046&12 24 24.92& 7 20 46.9&      114.4&\phn2.87$\pm$0.95&\phn3.03&\phn11.73&$-0.08^{+0.52}_{-0.48}$&$-0.27^{+0.65}_{-0.47}$&\cr
62&CXOU J122435.9$+$071925&12 24 35.93& 7 19 25.9&      116.3&\phn3.82$\pm$1.03&\phn3.72&\phn15.59&$+0.00^{+0.83}_{-0.83}$&$+0.69^{+0.25}_{-0.72}$&\cr
63&CXOU J122420.7$+$071938&12 24 20.72& 7 19 38.0&      117.2&\phn6.59$\pm$1.34&\phn4.93&\phn26.93&$+0.12^{+0.30}_{-0.33}$&$-0.11^{+0.38}_{-0.35}$&\cr
64&CXOU J122421.6$+$071757&12 24 21.62& 7 17 57.6&      118.8&\phn3.94$\pm$1.04&\phn3.80&\phn16.10&$+1.00^{+0.00}_{-0.48}$&$+1.00^{+0.00}_{-0.37}$&\cr
65&CXOU J122435.3$+$071805&12 24 35.30& 7 18 05.4&      119.9&   10.81$\pm$1.70&\phn6.34&\phn44.13&$-0.08^{+0.21}_{-0.20}$&$-0.71^{+0.39}_{-0.18}$&\cr
66&CXOU J122425.8$+$071706&12 24 25.81& 7 17 06.1&      123.2&\phn3.41$\pm$0.96&\phn3.54&\phn13.92&$-0.33^{+0.46}_{-0.34}$&$-0.66^{+0.74}_{-0.27}$&\cr
67&CXOU J122423.3$+$071720&12 24 23.35& 7 17 20.6&      126.4&\phn3.60$\pm$0.99&\phn3.62&\phn14.70&$+0.29^{+0.42}_{-0.57}$&$+0.17^{+0.49}_{-0.59}$&\cr
68&CXOU J122435.8$+$071806&12 24 35.88& 7 18 06.2&      127.2&\phn3.87$\pm$1.04&\phn3.73&\phn15.82&$+0.31^{+0.38}_{-0.51}$&$+0.08^{+0.51}_{-0.56}$&\cr
69&CXOU J122426.3$+$071653&12 24 26.39& 7 16 53.2&      133.5&   48.43$\pm$3.54&   13.67&   197.75&$+0.65^{+0.08}_{-0.10}$&$+0.64^{+0.09}_{-0.11}$&\cr
70&CXOU J122426.7$+$071650&12 24 26.70& 7 16 50.2&      135.6&\phn8.21$\pm$1.47&\phn5.58&\phn33.52&$+0.24^{+0.24}_{-0.27}$&$-0.16^{+0.35}_{-0.31}$&\cr
71&CXOU J122420.1$+$071739&12 24 20.18& 7 17 39.4&      146.8&   16.46$\pm$2.08&\phn7.89&\phn67.20&$+0.04^{+0.16}_{-0.16}$&$-0.44^{+0.23}_{-0.18}$&\cr
72&CXOU J122437.0$+$071757&12 24 37.08& 7 17 57.3&      147.1&\phn3.59$\pm$1.01&\phn3.57&\phn14.65&$+0.46^{+0.30}_{-0.46}$&$-0.29^{+0.77}_{-0.51}$&\cr
73&CXOU J122423.9$+$071651&12 24 23.95& 7 16 51.0&      147.4&\phn5.06$\pm$1.17&\phn4.34&\phn20.66&$+0.47^{+0.26}_{-0.38}$&$-0.33^{+0.72}_{-0.47}$&\cr
74&CXOU J122438.4$+$071902&12 24 38.46& 7 19 02.8&      151.8&\phn4.67$\pm$1.21&\phn3.86&\phn19.05&$-0.10^{+0.49}_{-0.45}$&$+0.15^{+0.39}_{-0.44}$&\cr
75&CXOU J122419.2$+$072019&12 24 19.25& 7 20 19.3&      153.9&\phn2.88$\pm$0.90&\phn3.20&\phn11.78&$+0.38^{+0.50}_{-0.87}$&$+0.55^{+0.36}_{-0.82}$&\cr
76&CXOU J122418.0$+$071839&12 24 18.01& 7 18 39.1&      154.4&\phn3.36$\pm$0.97&\phn3.46&\phn13.70&$-0.28^{+0.47}_{-0.37}$&$-1.00^{+2.00}_{-0.00}$&\cr
77&CXOU J122418.0$+$071803&12 24 18.04& 7 18 03.5&      163.6&\phn3.45$\pm$0.98&\phn3.52&\phn14.08&$+0.00^{+0.45}_{-0.45}$&$-0.11^{+0.51}_{-0.46}$&\cr
78&CXOU J122422.1$+$071646&12 24 22.19& 7 16 46.1&      164.7&\phn9.95$\pm$1.62&\phn6.14&\phn40.63&$+0.15^{+0.24}_{-0.26}$&$+0.10^{+0.26}_{-0.27}$&\cr
79&CXOU J122434.9$+$072116&12 24 34.93& 7 21 16.1&      165.4&\phn2.86$\pm$0.95&\phn3.02&\phn11.67&$+0.09^{+0.55}_{-0.61}$&$+0.17^{+0.50}_{-0.60}$&\cr
80&CXOU J122423.4$+$072136&12 24 23.44& 7 21 36.5&      168.7&\phn2.77$\pm$0.91&\phn3.03&\phn11.29&$+0.07^{+0.47}_{-0.50}$&$-0.37^{+0.82}_{-0.48}$&\cr
81&CXOU J122436.4$+$071706&12 24 36.47& 7 17 06.7&      169.2&\phn3.34$\pm$0.96&\phn3.49&\phn13.63&$-0.27^{+0.51}_{-0.39}$&$-0.56^{+0.66}_{-0.32}$&\cr
82&CXOU J122417.9$+$071741&12 24 17.91& 7 17 41.4&      174.7&\phn4.85$\pm$1.15&\phn4.23&\phn19.78&$+0.38^{+0.37}_{-0.56}$&$+0.51^{+0.29}_{-0.50}$&\cr
83&CXOU J122417.8$+$071715&12 24 17.80& 7 17 15.6&      189.5&\phn7.11$\pm$1.38&\phn5.17&\phn29.02&$+0.40^{+0.25}_{-0.33}$&$+0.03^{+0.41}_{-0.42}$&\cr
84&CXOU J122436.6$+$072129&12 24 36.65& 7 21 29.3&      191.8&\phn9.43$\pm$1.81&\phn5.22&\phn38.51&$-0.11^{+0.29}_{-0.27}$&$-0.12^{+0.35}_{-0.32}$&\cr
85&CXOU J122425.1$+$072211&12 24 25.19& 7 22 11.2&      192.9&\phn3.33$\pm$1.03&\phn3.24&\phn13.59&$+0.39^{+0.34}_{-0.49}$&$+0.18^{+0.46}_{-0.56}$&\cr
86&CXOU J122438.3$+$072107&12 24 38.31& 7 21 07.0&      193.7&   28.87$\pm$2.88&   10.01&   117.88&$-0.25^{+0.12}_{-0.11}$&$-0.61^{+0.15}_{-0.12}$&g\cr
87&CXOU J122427.2$+$072231&12 24 27.28& 7 22 31.2&      207.9&\phn8.82$\pm$1.87&\phn4.72&\phn36.03&$+0.32^{+0.34}_{-0.44}$&$+0.32^{+0.36}_{-0.49}$&\cr
88&CXOU J122414.7$+$072004&12 24 14.74& 7 20 04.9&      210.3&   37.47$\pm$3.20&   11.72&   153.01&$-0.22^{+0.11}_{-0.10}$&$-0.54^{+0.12}_{-0.10}$&d\cr
89&CXOU J122424.8$+$072229&12 24 24.83& 7 22 29.4&      211.9&\phn4.53$\pm$1.35&\phn3.36&\phn18.48&$+0.73^{+0.24}_{-1.10}$&$+0.76^{+0.22}_{-1.08}$&\cr
90&CXOU J122441.0$+$072047&12 24 41.05& 7 20 47.8&      216.9&\phn4.95$\pm$1.38&\phn3.59&\phn20.21&$+0.86^{+0.14}_{-1.67}$&$+0.85^{+0.15}_{-1.67}$&\cr
91&CXOU J122438.9$+$071625&12 24 38.96& 7 16 25.8&      224.3&   12.21$\pm$1.84&\phn6.65&\phn49.87&$+0.05^{+0.20}_{-0.20}$&$-0.50^{+0.32}_{-0.23}$&\cr
92&CXOU J122438.9$+$072152&12 24 38.93& 7 21 52.4&      231.6&\phn7.80$\pm$1.61&\phn4.83&\phn31.85&$+0.57^{+0.28}_{-0.52}$&$+0.68^{+0.21}_{-0.44}$&\cr
93&CXOU J122425.4$+$072258&12 24 25.42& 7 22 58.7&      238.6&\phn7.47$\pm$1.62&\phn4.60&\phn30.48&$+0.30^{+0.26}_{-0.32}$&$-0.33^{+0.66}_{-0.44}$&\cr
94&CXOU J122435.9$+$071533&12 24 35.94& 7 15 33.3&      239.6&\phn5.70$\pm$1.30&\phn4.39&\phn23.26&$+0.11^{+0.34}_{-0.37}$&$-0.05^{+0.42}_{-0.40}$&\cr
95&CXOU J122420.8$+$072237&12 24 20.84& 7 22 37.0&      240.1&\phn6.85$\pm$1.55&\phn4.41&\phn27.96&$+0.50^{+0.25}_{-0.38}$&$-0.10^{+0.62}_{-0.55}$&\cr
96&CXOU J122440.9$+$072204&12 24 40.90& 7 22 04.1&      260.6&   55.42$\pm$3.99&   13.88&   226.30&$-0.19^{+0.08}_{-0.08}$&$-0.48^{+0.10}_{-0.09}$&\cr
97&CXOU J122437.9$+$072311&12 24 37.91& 7 23 11.0&      285.9&   22.69$\pm$2.74&\phn8.27&\phn92.64&$-0.01^{+0.16}_{-0.16}$&$-0.25^{+0.21}_{-0.19}$&d\cr
98&CXOU J122419.9$+$072332&12 24 19.91& 7 23 32.6&      296.1&\phn3.83$\pm$1.27&\phn3.02&\phn15.64&$+0.08^{+0.53}_{-0.58}$&$+0.12^{+0.58}_{-0.67}$&d\cr
99&CXOU J122416.0$+$072324&12 24 16.03& 7 23 24.3&      317.7&   36.75$\pm$4.33&\phn8.65&   150.05&$+0.49^{+0.15}_{-0.18}$&$+0.43^{+0.17}_{-0.20}$&c\cr
\tableline
\end{tabular}}
\end{center}
\tablecomments{The units for $L_X$ are $10^{37}$ ergs s$^{-1}$ in the
0.3--10 keV band.}
\tablenotetext{a}{\scriptsize Sources near the center may be confused with
nearby sources, making their positions, fluxes, and extents uncertain.}
\tablenotetext{b}{\scriptsize Source is noticeably more extended than PSF.}
\tablenotetext{c}{\scriptsize Source is at the edge of the S3 detector, and
flux is uncertain due to large exposure correction.}
\tablenotetext{d}{\scriptsize Possible faint optical counterpart.}
\tablenotetext{e}{\scriptsize Identified with known AGN.}
\tablenotetext{f}{\scriptsize Globular cluster is possible optical counterpart.}
\tablenotetext{g}{\scriptsize Source may be variable.}
\end{table}

\clearpage
\begin{table}[p]
\begin{center}
\caption{Discrete X-ray Sources in NGC~4382 (M85)\label{tab:src_n4382}}
\tiny{
\begin{tabular}{lcccrrrrccl}
\tableline
\tableline
Src.&
&
R.A.&
Dec.&
\multicolumn{1}{c}{$d$}&
\multicolumn{1}{c}{Count Rate}&
&
&
&
&
\\
No.&
Name&
(h:m:s)&
($\arcdeg$:$\arcmin$:$\arcsec$)&
\multicolumn{1}{c}{($\arcsec$)}&
\multicolumn{1}{c}{($10^{-4}$ s$^{-1}$)}&
\multicolumn{1}{c}{SNR}&
\multicolumn{1}{c}{$L_X$}&
H21&
H31&
Notes \\
(1)&
(2)&
(3)&
(4)&
\multicolumn{1}{c}{(5)}&
\multicolumn{1}{c}{(6)}&
\multicolumn{1}{c}{(7)}&
\multicolumn{1}{c}{(8)}&
(9)&
(10)&
(11)
\\
\tableline
1 &CXOU J122524.0$+$181127&12 25 24.06&18 11 27.7&\phn\phn1.8&   \phn22.79$\pm$2.55&\phn8.93&\phn89.66&$-0.16^{+0.13}_{-0.13}$&$-0.36^{+0.14}_{-0.13}$&b\cr
2 &CXOU J122523.9$+$181121&12 25 23.99&18 11 21.7&\phn\phn4.2&   \phn16.44$\pm$2.15&\phn7.63&\phn64.71&$-0.10^{+0.17}_{-0.16}$&$-0.39^{+0.20}_{-0.17}$&\cr
3 &CXOU J122523.9$+$181131&12 25 23.90&18 11 31.1&\phn\phn5.5&   \phn10.90$\pm$1.79&\phn6.09&\phn42.88&$-0.08^{+0.25}_{-0.24}$&$-0.68^{+0.46}_{-0.21}$&\cr
4 &CXOU J122523.5$+$181126&12 25 23.58&18 11 26.7&\phn\phn6.6&   \phn11.87$\pm$1.88&\phn6.33&\phn46.71&$-0.04^{+0.27}_{-0.27}$&$-0.07^{+0.29}_{-0.27}$&\cr
5 &CXOU J122524.5$+$181134&12 25 24.54&18 11 34.9&   \phn11.5&   \phn11.97$\pm$1.84&\phn6.52&\phn47.11&$+0.35^{+0.24}_{-0.29}$&$+0.28^{+0.26}_{-0.30}$&\cr
6 &CXOU J122523.6$+$181136&12 25 23.60&18 11 36.4&   \phn12.2&\phn\phn5.10$\pm$1.24&\phn4.12&\phn20.06&$+0.26^{+0.37}_{-0.46}$&$+0.18^{+0.42}_{-0.49}$&\cr
7 &CXOU J122523.7$+$181139&12 25 23.75&18 11 39.9&   \phn14.6&\phn\phn9.64$\pm$1.66&\phn5.79&\phn37.92&$-0.20^{+0.20}_{-0.18}$&$-0.65^{+0.27}_{-0.17}$&\cr
8 &CXOU J122524.7$+$181141&12 25 24.70&18 11 41.0&   \phn17.7&   \phn11.31$\pm$1.79&\phn6.33&\phn44.48&$+0.69^{+0.17}_{-0.31}$&$+0.46^{+0.29}_{-0.43}$&\cr
9 &CXOU J122525.6$+$181141&12 25 25.65&18 11 41.5&   \phn27.7&\phn\phn5.24$\pm$1.25&\phn4.20&\phn20.63&$-0.09^{+0.45}_{-0.41}$&$+0.22^{+0.32}_{-0.37}$&\cr
10&CXOU J122522.8$+$181053&12 25 22.84&18 10 53.6&   \phn36.5&\phn\phn7.19$\pm$1.43&\phn5.01&\phn28.28&$-0.41^{+0.27}_{-0.21}$&$-0.97^{+1.90}_{-0.03}$&\cr
11&CXOU J122524.8$+$181048&12 25 24.83&18 10 48.7&   \phn38.9&\phn\phn3.32$\pm$0.99&\phn3.36&\phn13.08&$-0.36^{+0.63}_{-0.41}$&$-0.21^{+0.54}_{-0.43}$&\cr
12&CXOU J122524.9$+$181205&12 25 24.95&18 12 05.7&   \phn41.8&   \phn27.91$\pm$2.76&   10.12&   109.83&$+0.11^{+0.14}_{-0.15}$&$+0.00^{+0.15}_{-0.15}$&\cr
13&CXOU J122526.9$+$181149&12 25 26.94&18 11 49.7&   \phn47.7&   \phn24.71$\pm$2.59&\phn9.55&\phn97.24&$-0.03^{+0.14}_{-0.14}$&$-0.45^{+0.17}_{-0.14}$&\cr
14&CXOU J122520.5$+$181121&12 25 20.53&18 11 21.8&   \phn50.1&\phn\phn2.74$\pm$0.91&\phn3.01&\phn10.78&$-0.41^{+0.86}_{-0.47}$&$+0.14^{+0.45}_{-0.51}$&\cr
15&CXOU J122522.1$+$181043&12 25 22.15&18 10 43.5&   \phn50.3&   \phn17.38$\pm$2.22&\phn7.82&\phn68.40&$+0.21^{+0.18}_{-0.19}$&$-0.02^{+0.22}_{-0.22}$&\cr
16&CXOU J122520.6$+$181052&12 25 20.69&18 10 52.5&   \phn58.3&\phn\phn2.80$\pm$0.93&\phn3.01&\phn11.03&$-0.11^{+0.68}_{-0.59}$&$+0.17^{+0.49}_{-0.59}$&\cr
17&CXOU J122527.4$+$181157&12 25 27.47&18 11 57.8&   \phn58.4&   \phn10.08$\pm$1.67&\phn6.03&\phn39.65&$+0.22^{+0.24}_{-0.27}$&$+0.06^{+0.29}_{-0.30}$&\cr
18&CXOU J122519.7$+$181125&12 25 19.75&18 11 25.0&   \phn61.2&\phn\phn4.49$\pm$1.14&\phn3.92&\phn17.66&$-0.07^{+0.38}_{-0.36}$&$-0.53^{+0.71}_{-0.34}$&\cr
19&CXOU J122527.2$+$181215&12 25 27.29&18 12 15.4&   \phn67.8&\phn\phn7.24$\pm$1.56&\phn4.65&\phn28.48&$-0.14^{+0.32}_{-0.30}$&$-0.31^{+0.40}_{-0.31}$&\cr
20&CXOU J122521.7$+$181026&12 25 21.72&18 10 26.3&   \phn68.1&\phn\phn9.07$\pm$1.57&\phn5.79&\phn35.69&$+0.09^{+0.28}_{-0.29}$&$+0.09^{+0.28}_{-0.29}$&\cr
21&CXOU J122528.3$+$181205&12 25 28.35&18 12 05.7&   \phn73.2&\phn\phn6.11$\pm$1.40&\phn4.35&\phn24.04&$+1.00^{+0.00}_{-0.29}$&$+1.00^{+0.00}_{-0.26}$&\cr
22&CXOU J122521.6$+$181232&12 25 21.68&18 12 32.1&   \phn74.2&\phn\phn3.54$\pm$1.01&\phn3.51&\phn13.91&$-0.33^{+0.57}_{-0.40}$&$+0.05^{+0.39}_{-0.40}$&\cr
23&CXOU J122518.4$+$181137&12 25 18.42&18 11 37.0&   \phn80.9&   \phn14.54$\pm$1.98&\phn7.36&\phn57.22&$-0.20^{+0.22}_{-0.20}$&$-0.11^{+0.21}_{-0.20}$&\cr
24&CXOU J122519.8$+$181235&12 25 19.82&18 12 35.3&   \phn91.8&\phn\phn3.61$\pm$1.01&\phn3.59&\phn14.19&$+0.54^{+0.30}_{-0.55}$&$-0.07^{+0.73}_{-0.67}$&\cr
25&CXOU J122520.7$+$180955&12 25 20.72&18 09 55.7&      101.8&\phn\phn4.65$\pm$1.14&\phn4.09&\phn18.29&$-0.12^{+0.39}_{-0.36}$&$-0.35^{+0.53}_{-0.37}$&\cr
26&CXOU J122520.3$+$181301&12 25 20.32&18 13 01.7&      109.5&   \phn40.73$\pm$3.37&   12.10&   160.27&$+0.21^{+0.10}_{-0.10}$&$-0.60^{+0.15}_{-0.12}$&\cr
27&CXOU J122525.5$+$180936&12 25 25.51&18 09 36.8&      111.1&\phn\phn3.82$\pm$1.02&\phn3.73&\phn15.02&$+0.67^{+0.30}_{-1.09}$&$+0.82^{+0.16}_{-0.89}$&\cr
28&CXOU J122531.6$+$181051&12 25 31.68&18 10 51.8&      114.1&\phn\phn5.66$\pm$1.26&\phn4.49&\phn22.26&$-0.33^{+0.35}_{-0.28}$&$-0.47^{+0.41}_{-0.27}$&\cr
29&CXOU J122527.4$+$181314&12 25 27.48&18 13 14.1&      118.8&\phn\phn3.33$\pm$0.98&\phn3.38&\phn13.09&$-0.48^{+0.58}_{-0.34}$&$-0.48^{+0.63}_{-0.35}$&\cr
30&CXOU J122521.6$+$180929&12 25 21.61&18 09 29.0&      121.9&   \phn21.55$\pm$2.39&\phn9.01&\phn84.78&$+0.10^{+0.16}_{-0.16}$&$+0.02^{+0.17}_{-0.17}$&\cr
31&CXOU J122518.1$+$181255&12 25 18.19&18 12 55.7&      122.5&\phn\phn2.68$\pm$0.87&\phn3.06&\phn10.55&$-0.41^{+0.88}_{-0.47}$&$-0.04^{+0.57}_{-0.54}$&\cr
32&CXOU J122516.8$+$181234&12 25 16.82&18 12 34.3&      123.5&\phn\phn8.37$\pm$1.51&\phn5.53&\phn32.93&$+0.22^{+0.23}_{-0.25}$&$-0.57^{+0.46}_{-0.26}$&\cr
33&CXOU J122530.4$+$181311&12 25 30.41&18 13 11.8&      139.5&   \phn12.51$\pm$1.88&\phn6.65&\phn49.24&$+0.47^{+0.18}_{-0.23}$&$+0.20^{+0.27}_{-0.30}$&\cr
34&CXOU J122522.2$+$181344&12 25 22.22&18 13 44.0&      140.5&\phn\phn7.37$\pm$1.46&\phn5.04&\phn28.99&$+0.55^{+0.24}_{-0.38}$&$+0.47^{+0.28}_{-0.42}$&\cr
35&CXOU J122515.0$+$181018&12 25 15.01&18 10 18.8&      145.1&\phn\phn4.20$\pm$1.11&\phn3.77&\phn16.52&$+0.71^{+0.25}_{-0.79}$&$+0.41^{+0.49}_{-0.95}$&\cr
36&CXOU J122513.8$+$181248&12 25 13.80&18 12 48.9&      167.9&\phn\phn2.70$\pm$0.88&\phn3.05&\phn10.62&$-0.77^{+0.98}_{-0.21}$&$-0.35^{+0.50}_{-0.36}$&\cr
37&CXOU J122517.6$+$181350&12 25 17.64&18 13 50.2&      170.7&   \phn22.95$\pm$2.61&\phn8.79&\phn90.29&$-0.05^{+0.16}_{-0.16}$&$-0.10^{+0.17}_{-0.17}$&\cr
38&CXOU J122517.1$+$181346&12 25 17.17&18 13 46.8&      171.6&      156.44$\pm$7.25&   21.60&   615.56&$-0.19^{+0.05}_{-0.05}$&$-0.59^{+0.06}_{-0.05}$&d,e\cr
39&CXOU J122512.9$+$181016&12 25 12.98&18 10 16.2&      172.3&   \phn16.00$\pm$2.06&\phn7.76&\phn62.94&$+0.00^{+0.20}_{-0.20}$&$-0.05^{+0.20}_{-0.20}$&\cr
40&CXOU J122528.0$+$181417&12 25 28.06&18 14 17.2&      180.7&\phn\phn3.56$\pm$1.04&\phn3.42&\phn14.08&$-0.15^{+0.60}_{-0.51}$&$-0.15^{+0.56}_{-0.47}$&\cr
41&CXOU J122518.1$+$180841&12 25 18.12&18 08 41.5&      184.8&\phn\phn8.19$\pm$2.66&\phn3.09&\phn32.24&$-0.14^{+0.60}_{-0.51}$&$-0.14^{+0.60}_{-0.51}$&c\cr
42&CXOU J122536.5$+$181002&12 25 36.58&18 10 02.6&      197.2&\phn\phn6.43$\pm$1.37&\phn4.70&\phn25.31&$-0.23^{+0.31}_{-0.27}$&$-0.49^{+0.43}_{-0.28}$&\cr
43&CXOU J122511.1$+$181006&12 25 11.18&18 10 06.1&      199.9&\phn\phn2.70$\pm$0.89&\phn3.05&\phn10.64&$+0.47^{+0.35}_{-0.59}$&$-0.45^{+1.23}_{-0.51}$&\cr
44&CXOU J122528.6$+$180808&12 25 28.66&18 08 08.6&      208.0&\phn\phn3.20$\pm$0.94&\phn3.39&\phn12.59&$-0.36^{+0.63}_{-0.41}$&$-0.26^{+0.57}_{-0.44}$&\cr
45&CXOU J122523.3$+$180758&12 25 23.33&18 07 58.1&      208.0&   \phn22.72$\pm$4.02&\phn5.66&\phn89.40&$+0.00^{+0.24}_{-0.24}$&$-0.50^{+0.36}_{-0.24}$&c\cr
46&CXOU J122538.8$+$181137&12 25 38.83&18 11 37.1&      211.1&\phn\phn5.71$\pm$1.41&\phn4.06&\phn22.48&$+1.00^{+0.00}_{-0.40}$&$+1.00^{+0.00}_{-0.09}$&\cr
47&CXOU J122522.4$+$181457&12 25 22.48&18 14 57.1&      212.3&   \phn15.34$\pm$2.27&\phn6.75&\phn60.36&$-0.24^{+0.22}_{-0.20}$&$-0.43^{+0.30}_{-0.23}$&\cr
48&CXOU J122538.7$+$181205&12 25 38.75&18 12 05.1&      213.2&\phn\phn6.64$\pm$1.53&\phn4.33&\phn26.13&$+0.66^{+0.26}_{-0.68}$&$+0.67^{+0.25}_{-0.67}$&d\cr
49&CXOU J122512.8$+$181409&12 25 12.80&18 14 09.3&      228.8&   \phn20.54$\pm$2.72&\phn7.55&\phn80.83&$+0.10^{+0.17}_{-0.18}$&$-0.35^{+0.25}_{-0.21}$&\cr
50&CXOU J122508.0$+$181229&12 25 08.00&18 12 29.9&      237.3&\phn\phn3.75$\pm$1.11&\phn3.37&\phn14.73&$+0.30^{+0.34}_{-0.43}$&$-0.65^{+1.25}_{-0.33}$&\cr
51&CXOU J122540.6$+$181224&12 25 40.60&18 12 24.6&      243.2&   \phn10.99$\pm$1.94&\phn5.67&\phn43.24&$+0.13^{+0.24}_{-0.25}$&$-0.08^{+0.33}_{-0.32}$&\cr
52&CXOU J122509.2$+$181336&12 25 09.27&18 13 36.7&      247.8&\phn\phn9.67$\pm$2.13&\phn4.55&\phn38.07&$+0.36^{+0.31}_{-0.41}$&$+0.28^{+0.35}_{-0.45}$&c\cr
53&CXOU J122506.1$+$181045&12 25 06.15&18 10 45.8&      258.0&   \phn12.67$\pm$1.95&\phn6.49&\phn49.83&$+0.11^{+0.22}_{-0.23}$&$-0.19^{+0.30}_{-0.27}$&d\cr
54&CXOU J122505.9$+$181158&12 25 05.91&18 11 58.9&      260.5&\phn\phn9.25$\pm$1.67&\phn5.53&\phn36.41&$+0.12^{+0.25}_{-0.27}$&$-0.48^{+0.57}_{-0.33}$&g\cr
55&CXOU J122504.7$+$181133&12 25 04.74&18 11 33.5&      275.1&   \phn13.21$\pm$1.98&\phn6.66&\phn51.98&$+0.58^{+0.20}_{-0.30}$&$+0.59^{+0.19}_{-0.29}$&\cr
56&CXOU J122543.4$+$181048&12 25 43.44&18 10 48.3&      278.9&\phn\phn8.82$\pm$2.98&\phn3.02&\phn34.69&$+0.45^{+0.36}_{-0.60}$&$-0.14^{+0.88}_{-0.70}$&c\cr
57&CXOU J122503.7$+$181124&12 25 03.70&18 11 24.1&      289.9&   \phn14.61$\pm$2.52&\phn5.82&\phn57.48&$-0.12^{+0.23}_{-0.22}$&$-0.49^{+0.31}_{-0.22}$&c,d\cr
58&CXOU J122546.0$+$181302&12 25 46.05&18 13 02.9&      328.3&   \phn12.11$\pm$2.49&\phn4.86&\phn47.65&$+0.24^{+0.32}_{-0.38}$&$-0.40^{+1.03}_{-0.52}$&\cr
\tableline
\end{tabular}}
\end{center}
\tablecomments{The notation is the same as in
Table~\protect\ref{tab:src_n4365}.}
\end{table}

\clearpage

\begin{deluxetable}{llcclccccccrc}
\tabletypesize{\scriptsize}
\rotate
\tablewidth{0pt}
\tablecaption{X-ray Spectral Fits of NGC~4365 \label{tab:spectra_n4365}}
\tablehead{
&&&&
\multicolumn{3}{c}{Hard Component}&&
\multicolumn{3}{c}{Soft Component (mekal)}&&\\
\cline{5-7} \cline{9-11}
&&&
\colhead{$N_H$}&
\colhead{Model}&
\colhead{$kT_h$ or $\Gamma$}&
\colhead{$F^h_X$}&&
\colhead{$kT_s$}&
\colhead{Abund.}&
\colhead{$F^s_X$)}&&\\
\colhead{Row}&
\colhead{Origin}&
\colhead{Region}&
\colhead{($10^{20}$ cm$^{-2}$)}&&
\colhead{(keV)}&
\colhead{(\tablenotemark{a} \,)}&&
\colhead{(keV)}&
\colhead{(solar)}&
\colhead{(\tablenotemark{a} \,)}&
\colhead{Counts}&
\colhead{$\chi^2$/dof}}
\startdata
1                 &Sources   &$<$ 1 $a_{\rm eff}$&(1.63)                    &Bremss&$6.08^{+2.71}_{-1.62}$&2.06&&&&(0.0)& 901&\phn\phn41.7/33=1.26\\
2                 &Sources   &$<$ 1 $a_{\rm eff}$&0.00 [$<$3.66]            &Bremss&$6.57^{+3.11}_{-1.83}$&2.05&&&&(0.0)& 901&\phn\phn40.6/32=1.27\\
3\tablenotemark{b}&Sources   &$<$ 1 $a_{\rm eff}$&(1.63)                    &Power &$1.67^{+0.12}_{-0.12}$&2.35&&&&(0.0)& 901&\phn\phn38.6/33=1.17\\
4                 &Sources   &$<$ 1 $a_{\rm eff}$&0.00 [$<$9.74]            &Power &$1.64^{+0.23}_{-0.11}$&2.34&&&&(0.0)& 901&\phn\phn38.4/32=1.20\\
5                 &Sources   &$1-3$ $a_{\rm eff}$&(1.63)                    &Power &(1.67)                &2.34&&&&(0.0)& 887&\phn\phn38.9/33=1.18\\
6                 &Sources   &$1-3$ $a_{\rm eff}$&(1.63)                    &Power &$1.56^{+0.11}_{-0.11}$&2.47&&&&(0.0)& 887&\phn\phn36.0/32=1.13\\
7                 &Sources   &$1-3$ $a_{\rm eff}$&$10.34^{+10.94}_{-10.28}$ &Power &$1.73^{+0.24}_{-0.22}$&2.58&&&&(0.0)& 887&\phn\phn34.1/31=1.10\\
8                 &Sources   &Field              &(1.63)                    &Power &(1.67)                &6.94&&&&(0.0)&2660&\phn\phn85.6/89=0.96\\
9                 &Sources   &Field              &(1.63)                    &Power &$1.58^{+0.07}_{-0.06}$&7.28&&&&(0.0)&2660&\phn\phn79.8/88=0.91\\
10                &Sources   &Field              &3.04 [$<$8.76]            &Power &$1.61^{+0.14}_{-0.11}$&7.32&&&&(0.0)&2660&\phn\phn79.6/87=0.92\\
11\tablenotemark{c}  &Unresolved&$<$ 1 $a_{\rm eff}$&(1.63)                  &      &                        &(0.0)         &&$0.60^{+0.06}_{-0.07}$&$0.03^{+0.02}_{-0.02}$&1.59& 935&\phn\phn71.3/48=1.49\\
12\tablenotemark{c}  &Unresolved&$<$ 1 $a_{\rm eff}$&0.00 [$<$13.11]         &      &                        &(0.0)         &&$0.61^{+0.06}_{-0.07}$&$0.03^{+0.03}_{-0.02}$&1.53& 935&\phn\phn71.1/47=1.51\\
13\tablenotemark{b,c}&Unresolved&$<$ 1 $a_{\rm eff}$&(1.63)                  &Power &(1.67)                  & 1.30         &&$0.56^{+0.05}_{-0.08}$&0.35 [$>$0.08]        &0.68& 935&\phn\phn34.4/47=0.73\\
14\tablenotemark{c}  &Unresolved&$1-3$ $a_{\rm eff}$&(1.63)                  &Power &(1.67)                  & 1.51         &&(0.56)                &(0.35)                &0.84&1155&      195.2/170=1.15\\
15\tablenotemark{c}  &Unresolved&$1-3$ $a_{\rm eff}$&(1.63)                  &Power &(1.67)                  &0.00 [$<$0.34]&&$0.80^{+0.09}_{-0.09}$&$0.07^{+0.05}_{-0.03}$&1.77&1155&      158.6/168=0.94\\
16\tablenotemark{b,c}  &Total     &$<$ 1 $a_{\rm eff}$&(1.63)                  &Power &(1.67)                  &3.66          &&(0.56)                &(0.35)                &0.75&1831&\phn\phn70.7/72=0.98\\
17\tablenotemark{c}  &Total     &$<$ 1 $a_{\rm eff}$&(1.63)                  &Power &(1.67)                  &3.56          &&$0.46^{+0.10}_{-0.23}$&0.09 [$>$0.00]        &1.09&1831&\phn\phn67.1/70=0.96\\
18\tablenotemark{c}  &Total     &$<$ 1 $a_{\rm eff}$&(1.63)                  &Power &$1.58^{+0.36}_{-0.39}$  &3.49          &&$0.46^{+0.10}_{-0.23}$&0.05 [$>$0.00]        &1.34&1831&\phn\phn67.0/69=0.97\\
19\tablenotemark{c}  &Total     &$<$ 1 $a_{\rm eff}$&(1.63)                  &Bremss&12.22 [$>5.21$]          &3.13         &&$0.46^{+0.10}_{-0.21}$&0.04 [$<$0.26]        &1.64&1831&\phn\phn66.6/69=0.96\\
20\tablenotemark{c}  &Total     &$<$ 3 $a_{\rm eff}$&(1.63)                  &Power &$1.69^{+0.65}_{-0.71}$  &5.55          &&$0.63^{+0.08}_{-0.09}$&0.03 [$>$0.01]        &3.32&3820&      199.0/220=0.90\\
21\tablenotemark{c}  &Total     &$<$ 3 $a_{\rm eff}$&(1.63)                  &Bremss&9.63 [$>$3.36]          &4.74          &&$0.63^{+0.08}_{-0.09}$&$0.03^{+0.06}_{-0.02}$&3.90&3820&      198.8/220=0.90\\

\enddata
\tablenotetext{a}{Units are $10^{-13}$ ergs cm$^{-2}$ s$^{-1}$ in 0.3--10
keV band.}
\tablenotetext{b}{The adopted best-fit model for this emission.}
\tablenotetext{c}{The energy range for this spectra excludes 1.6 - 1.9 keV.}
\end{deluxetable}

\clearpage
\begin{deluxetable}{llcclccccccrc}
\tabletypesize{\scriptsize}
\rotate
\tablewidth{0pt}
\tablecaption{X-ray Spectral Fits of NGC~4382 \label{tab:spectra_n4382}}
\tablehead{
&&&&
\multicolumn{3}{c}{Hard Component}&&
\multicolumn{3}{c}{Soft Component (mekal)}&&\\
\cline{5-7} \cline{9-11}
&&&
\colhead{$N_H$}&
\colhead{Model}&
\colhead{$kT_h$ or $\Gamma$}&
\colhead{$F^h_X$}&&
\colhead{$kT_s$}&
\colhead{Abund.}&
\colhead{$F^s_X$)}&&\\
\colhead{Row}&
\colhead{Origin}&
\colhead{Region}&
\colhead{($10^{20}$ cm$^{-2}$)}&&
\colhead{(keV)}&
\colhead{(\tablenotemark{a} \,)}&&
\colhead{(keV)}&
\colhead{(solar)}&
\colhead{(\tablenotemark{a} \,)}&
\colhead{Counts}&
\colhead{$\chi^2$/dof}}
\startdata
1                 &Sources   &$<$ 2 $a_{\rm eff}$&(2.45)                    &Bremss&$\phn9.70^{+\phn7.54}_{-3.20}$&2.47&&&&(0.0)& 921&\phn\phn30.3/33=0.92\\
2                 &Sources   &$<$ 2 $a_{\rm eff}$&0.00 [$<$6.01]            &Bremss&$11.48^{+10.77}_{-4.73}$      &2.47&&&&(0.0)& 921&\phn\phn29.5/32=0.92\\
3\tablenotemark{b}&Sources   &$<$ 2 $a_{\rm eff}$&(2.45)                    &Power &$1.52^{+0.11}_{-0.11}$        &2.73&&&&(0.0)& 921&\phn\phn29.2/33=0.89\\
4                 &Sources   &$<$ 2 $a_{\rm eff}$&2.43 [$<$12.63]           &Power &$1.52^{+0.23}_{-0.15}$        &2.73&&&&(0.0)& 921&\phn\phn29.2/32=0.91\\
5                 &Sources   &Field              &(2.45)                    &Power &(1.52)                        &6.08&&&&(0.0)&2055&\phn\phn51.4/71=0.72\\
6                 &Sources   &Field              &(2.45)                    &Power &$1.57^{+0.08}_{-0.08}$        &5.91&&&&(0.0)&2055&\phn\phn50.0/70=0.71\\
7                 &Sources   &Field              &0.00 [$<$4.73]            &Power &$1.52^{+0.12}_{-0.07}$        &5.87&&&&(0.0)&2055&\phn\phn48.9/69=0.71\\
8\tablenotemark{c}   &Unresolved&$<$ 2 $a_{\rm eff}$&(2.45)                  &      &                              &    &&$0.31^{+0.05}_{-0.04}$&$0.08^{+0.08}_{-0.04}$  &5.68 &2620&150.4/129=1.17\\
9\tablenotemark{c}   &Unresolved&$<$ 2 $a_{\rm eff}$&0.00 [$<$4.38]          &      &                              &    &&$0.33^{+0.07}_{-0.05}$&$0.08^{+0.09}_{-0.04}$  &4.99 &2620&148.9/128=1.16\\
10\tablenotemark{c}  &Unresolved&$<$ 2 $a_{\rm eff}$&(2.45)                  &Power &(1.52)                        &1.57&&$0.29^{+0.04}_{-0.03}$&7.73 [$>$0.13]          &4.12 &2620&122.7/128=0.96\\
11\tablenotemark{b,c}&Unresolved&$<$ 2 $a_{\rm eff}$&(2.45)                  &Power &(1.52)                        &1.47&&$0.28^{+0.04}_{-0.03}$&(0.50)                  &4.61 &2620&123.0/129=0.95\\
12\tablenotemark{c}  &Unresolved&$1-3$ $a_{\rm eff}$&(2.45)                  &Power &(1.52)                        &0.89&&(0.28)                &(0.50)                  &3.31 &1966&193.8/195=0.99\\
13\tablenotemark{c}  &Unresolved&$1-3$ $a_{\rm eff}$&(2.45)                  &Power &(1.52)                        &0.80&&$0.30^{+0.07}_{-0.04}$&(0.50)                  &2.98 &1966&193.2/194=1.00\\
14\tablenotemark{b,c}&Total     &$<$ 2 $a_{\rm eff}$&(2.45)                  &Power &(1.52)                        &4.32&&(0.28)                &(0.50)                  &4.73 &3482&139.4/146=0.95\\
15\tablenotemark{c}  &Total     &$<$ 2 $a_{\rm eff}$&(2.45)                  &Power &(1.52)                        &4.19&&$0.31^{+0.05}_{-0.04}$&(0.50)                  &4.14 &3482&138.1/145=0.95\\
16\tablenotemark{c}  &Total     &$<$ 2 $a_{\rm eff}$&(2.45)                  &Power &$1.31^{+0.24}_{-0.24}$        &4.53&&$0.33^{+0.06}_{-0.04}$&(0.50)                  &4.02 &3482&136.0/144=0.94\\
17\tablenotemark{c}  &Total     &$<$ 2 $a_{\rm eff}$&(2.45)                  &Brem  &45.00 [$>10.38$]              &4.39&&$0.33^{+0.06}_{-0.04}$&(0.50)                  &4.01 &3482&136.2/144=0.95\\
18\tablenotemark{c}  &Total     &$<$ 3 $a_{\rm eff}$&(2.45)                  &Power &$1.27^{+0.26}_{-0.25}$        &5.69&&$0.33^{+0.06}_{-0.04}$&(0.50)                  &5.06 &4452&245.0/237=1.03\\
19\tablenotemark{c}  &Total     &$<$ 3 $a_{\rm eff}$&(2.45)                  &Brem  &58.57 [$>10.47$]              &5.53&&$0.33^{+0.06}_{-0.04}$&(0.50)                  &5.08 &4452&245.0/237=1.03\\
\enddata   
\tablecomments{Notation is the same as Table~\protect\ref{tab:spectra_n4365}.}
\end{deluxetable}

\clearpage

\clearpage

\clearpage

\clearpage

\clearpage

\clearpage

\begin{figure}
\epsfig{file=f9a.eps,angle=-90,width=0.5\textwidth,clip=}
\epsfig{file=f9b.eps,angle=-90,width=0.5\textwidth,clip=}
\caption{
In the upper panels, the cumulative X-ray spectra of all of the resolved
sources in the inner effective radius of NGC~4365 (left) and inner two
effective radii of NGC~4382 (right) with $1 \sigma$ error bars
are overlayed by the solid
histograms of the best-fit model spectra (Table~\ref{tab:spectra_n4365} and
\ref{tab:spectra_n4382} row 3). Lower panels display the contribution to
$\chi^2$ with the sign indicating the sign of the residual.
\label{fig:src_spec}}
\end{figure}

\end{document}